\newcommand\nodata{{...} }

\newcommand\etal{et~al.}
\newcommand\kms{\ifmmode {\rm\,km\,s^{-1}}\else${\rm\,km\,s^{-1}}$\fi}
\font\aipsfont = cmsy9 scaled\magstep1

\newcommand\aips {{\aipsfont AIPS$\;$}}
\def\spose#1{\hbox to 0pt{#1\hss}}
\newcommand\simlt{\mathrel{\spose{\lower 3pt\hbox{$\mathchar"218$}}
     \raise 2.0pt\hbox{$\mathchar"13C$}}}
\newcommand\simgt{\mathrel{\spose{\lower 3pt\hbox{$\mathchar"218$}}
     \raise 2.0pt\hbox{$\mathchar"13E$}}}
\newcommand\aap{{\em A\&A}}
\newcommand\aaps{{\em A\&AS}}
\newcommand\aj{{\em AJ}}
\newcommand\apj{{\em ApJ}}
\newcommand\apjl{{\em ApJ}}

\newcommand\mnras{{\em MNRAS}}
\newcommand\nat{{\em Nature}}

\newcommand\pasp{{\em PASP}}
 
 \newcommand\Lya{Ly$\alpha$}

\newcommand\CIV{\hbox{C~IV}~$\lambda$1549}
\newcommand\HeII{\hbox{He~II}~$\lambda$1640}

\newcommand\CIII{\hbox{C~III]}~$\lambda$1909} \newcommand\CII{\hbox{C~$\rm II$]}~$\lambda$2326} \newcommand\NeIV{\hbox{[Ne~$\rm IV$}]~$\lambda$2424}
\newcommand\MgII{\hbox{Mg~$\rm II$}~$\lambda$2800}

\newcommand\OII{\hbox{[O~II}]~$\lambda$3727}

\newcommand\OIII{\hbox{[O~III}]~$\lambda$5007}

\documentclass[usenatbib]{mn2e}

\usepackage[dvips]{graphicx}
\usepackage{rotating}

\begin{document}

\title[Imaging and Spectroscopy of USS Radio Sources]{Imaging and Spectroscopy of Ultra Steep Spectrum Radio Sources
\thanks{Based on observations obtained at Cerro Tololo Inter-American Observatory, a division of the National Optical Astronomy Observatories, which is
operated by the Association of Universities for Research in Astronomy,
Inc.  under cooperative agreement with the National Science
Foundation. Based on observations with the Very Large Telescope,
obtained at the European Southern Observatory in Chile under Proposals
65.O-0125(A), 66.A.0006(B) and 69.A-0337(A).  }.}
\author[C. Bornancini \etal]{
\parbox[t]{\textwidth}{
Carlos~G.~Bornancini$^{1,2}$, Carlos De Breuck$^3$, Wim de Vries$^{4,5}$, Steve Croft$^{4,6}$, Wil van Breugel$^{4,6}$, Huub R\"ottgering$^7$ and Dante Minniti$^8$}
\vspace*{6pt} \\ 
$^1$Grupo de Investigaciones en Astronom\'\i a Te\'orica y Experimental, IATE, Observatorio Astron\'omico, Universidad Nacional de C\'ordoba,\\ 
Laprida 854, X5000BGR, C\'ordoba, Argentina.\\
$^2$ Secretar\'ia de Ciencia y T\'ecnica de la Universidad Nacional de C\'ordoba.\\
$^3$ European Southern Observatory, Karl Schwarzschild Stra\ss e 2, D-85748 Garching, Germany.\\
$^4$ Institute of Geophysics and Planetary Physics, Lawrence Livermore National Laboratory L-413, 7000 East Avenue, Livermore, CA 94550, USA \\
$^5$ University of California, Davis, 1 Shields Avenue, Davis, CA 95616, USA \\
$^6$ University of California, Merced, P.O. Box 2039, Merced, CA 95344, USA \\
$^7$ Leiden Observatory, PO Box 9513, 2300 RA, Leiden, The Netherlands\\
$^8$ Pontificia Universidad Cat\'olica de Chile, Departamento de Astronom\'\i a y Astrof\'\i sica, Casilla 306, Santiago 22, Chile }

\pubyear{2007}

\maketitle

\begin{abstract}

We present a sample of 40 Ultra Steep Spectrum (USS, $\alpha \leq -1.3$, $S_{\nu}\propto \nu^{\alpha}$) radio sources selected from the Westerbork in the Southern Hemisphere (WISH) catalog.
The USS sources have been imaged in $K$--band at the Cerro Tololo Inter-American Observatory (CTIO) and with the Very Large Telescope at Cerro Paranal. 
We also present VLT, Keck and Willian Herschel Telescope(WHT) optical
spectroscopy of 14 targets selection from 4 different USS samples. For 12
sources, we have been able to determine the redshifts, including 4 new
radio galaxies at $z>3$.
We find that most of our USS sources have predominantly small ($<$6\arcsec) radio sizes and faint magnitudes ($K$$\simgt$18).
The mean $K-$band counterpart magnitude is $\overline{K}$=18.6. 
The expected redshift distribution estimated using the Hubble $K-z$ diagram has a mean of $\overline{z}_{exp}$$\sim$2.13, which is higher than the predicted redshift obtained for the SUMSS--NVSS sample and the expected redshift obtained in the 6C$^{**}$ survey. 
The compact USS sample analyzed here may contain a higher fraction of galaxies which are high redshift and/or are heavily obscured by dust.
Using the 74, 352 and 1400\,MHz flux densities of a sub-sample, we construct a radio colour-colour diagram. We find that all but one of our USS sources have a strong tendency to flatten below 352\,MHz. We also find that the highest redshift source from this paper (at $z$=3.84) does not show evidence for spectral flattening down to 151\,MHz. This suggests that very low frequency selected USS samples will likely be more efficient to find high redshift galaxies.

\end{abstract}


\section{Introduction}
Distant powerful radio sources represent excellent targets to study evolutionary processes related to massive galaxies and their associated surrounding structures at high redshifts \citep[e.g.][]{DB02,mileynat,overzier,villar03,nesvadba}. 
 The most effective method for finding high-redshift radio galaxies is by selecting for Ultra Steep-Spectrum radio sources (USS, $\alpha \leq -1$, $S_{\nu}\propto \nu^{\alpha}$). A possible explanation of this empirical method is a $K$--correction induced by a curvature of the radio spectra. High redshift sources with concave radio spectra will have steeper spectral indices compared to those at low redshift \citep{krolik,carilli,gopal}. 
Comparing the extremely steep spectral index sources associated with galaxies residing closest to the cluster centres, \citet{klamer} found an explanation that suggests that the steeper spectra can also be explained by pressure-confined radio lobes which have slow adiabatic expansion losses in high density environments.
High-redshift radio galaxies are extremely luminous, and spatially extended, compared to normal galaxies at similar redshifts. 
The tight $K-z$ relation of radio galaxies in the Hubble $K$ diagram suggest that these objects are amongst the most massive systems at each redshift \citep{DB02,jarvis01,willot03,eales,lacy}.
Using the $K-z$ diagram and galaxy evolution models, \citet{rocca} found that the typical hosts of radio galaxies correspond to the most massive elliptical galaxies with baryon masses $M\sim10^{12}$$M_{\sun}$.

There is increasing new evidence that distant radio galaxies represent massive forming systems.
\citet{miley} found more than 10 individual clumpy features, possible satellite galaxies in the process of merging, in a region of 50$\times$40\,kpc around the radio galaxy MRC 1138--262 at $z=2.2$.
From a large VLT program, searching Ly$\alpha$ emitters around a sample of $2<z<5.2$ radio galaxies, \citet{vene} found that at least six of eight fields studied are overdense in Ly$\alpha$ emitters by a factor of 3-5 compared to the field.
Some radio-galaxies, selected with the USS criterion, are surrounded by giant Ly$\alpha$ haloes. The size of these gas structures (100--200\,kpc) are similar to the observed size of cD haloes in galaxies in nearby clusters \citep{reulanda, breugel06}. 
From a spatial cross-correlation analysis between USS at $0.5<z<1.5$ and surrounding $K<20$ galaxies, \citet{bornan} found a comoving correlation length comparable to those obtained for clusters of galaxies with masses in the range $M\sim10^{14} M_{\sun}$  in a cosmological N-body simulation. 
In this paper we present $K$--band observations, high-resolution radio maps and optical spectroscopy for a subsample
 of southern Ultra Steep Spectrum Radio Sources selected from the Westerbork in the Southern Hemisphere (WISH, \citet{wish}) 352\,MHz catalog. 
We also present optical spectroscopy of sources selected from the \citet{DB00} and WISH USS samples, yielding 12 new redshifts,
including four new radio galaxies at $z$$>$3
These provide excellent targets for follow-up observations in this interesting redshift range.

The structure of this paper is organized as follows: Section 2
describes the sample analyzed.  We describe the near-IR observations
and data reductions in Section 3.  Section 4 presents the near-IR
identification and source extraction.  We describe optical
spectroscopy observations in Section 5 and discuss our results in
Section 6.  Finally, Section 7 presents the main conclusions.

In this work we assume a standard $\Lambda$CDM model Universe with cosmological parameters, $\Omega_{M}$=0.3, $\Omega_{\Lambda}$=0.7 and a Hubble constant of $H_0=$100 Km~s$^{-1}$Mpc$^{-1}$.

\section{Sample definition}

The USS sample was selected from the 352 MHz Westerbork In the Southern Hemisphere catalogue \citep{wish}.
We used a selection criterion based on the radio properties of the sources, steep radio spectral index ($\alpha_{352}^{1400}<-1.3$) and small angular size ($<30\arcsec$, as measured on the VLA maps). All of the first and most of the second priority targets have been observed.
A detailed description of the radio observations and data reduction can be found in \citet{wish}. Most of the targets selected for optical spectroscopy are selected from the USS samples of \citet{DB00}.

\section{Observations and data reduction}

\subsection{Radio Imaging}
\subsubsection{VLA}
In order to obtain high resolution radio maps and accurate positions, we used the Very Large Array (VLA) of the National Radio Astronomy
Observatory in the hybrid BnA-configuration at 1.4\,GHz. The images were obtained on 1999 October 16 and 20 and 
consisted of L-band short subscans of 3 minutes each.
 We used the standard data
reduction recipes in the Astronomical Image Processing Software (\aips) package, including self-calibration for phase corrections.

\subsection{K-band Imaging}
\subsubsection{CTIO}
The USS fields were observed during two runs in 2000 March and
2001 January using the Ohio State InfraRed Imager/Spectrometer (OSIRIS) imager on the 4$-$meter V.M. Blanco 
telescope at the Cerro Tololo Inter-American Observatory (CTIO). OSIRIS is a 0.9--2.4 $\mu$m camera with a 1024x1024 HAWAII HgCdTe CCD array.
We used a focal ratio of f/7 which resulted in a respective pixel size of 0\farcs161.
Individual frames were obtained as a co-addition of a number of single exposures with different integration times, in this way one effectively removes row/column related defects.
For example, for sources with total exposures times of 1920s, we use 
12 single exposures of 10s observed with a 
a 16$-$point dithering pattern.  This results in $16\times10\times12=1920s=32$ minutes on source. 
We used similar procedure for the other images, doing small changes in the number of points of the dithering pattern and the exposure times for single observations 
During the observations, tip-tilt mirror corrections were made. 
This resulted in typical FWHMs on the individual pointing frames 
($12\times10s$ co-added) of $0\farcs5$ to $0\farcs7$. The mean was 
around $0\farcs6$ for the nights 2000 March 20 to 22, and around $0\farcs7$  on the 2001 January run. 

A subsample of USS sources was observed in 1999 July with the Cerro Tololo IR imager (CIRIM) camera at the CTIO 4$-$meter V.M. Blanco 
telescope. The CCD detector is a 256$\times$256 HgCdTe NICMOS 3 array with a pixel scale of 0\farcs414 using a focal ratio of f/7.5. 
For these observations we obtained a mean FWHM of $1\farcs1$.

\subsubsection{VLT}
Seven sources undetected in the CTIO images were observed with 
the Infrared Spectrometer And Array
Camera \citep[ISAAC; ][]{moor} on the Very Large Telescope (VLT) at Paranal, (Chile)
between 2002 April and September. A $K_{s}$ filter (2.0-2.3 $\micron$) was used.The pixel scale was 0\farcs148 per pixel .
All the images were obtained under optimum seeing conditions with FWHM in the range 0\farcs7 and 1\farcs0, and a number of standard stars were observed during the same nights.
In Table 1 we present the log of the near-IR observations, with the observation dates, telescope/instrument used and the total exposure times for all the USS sample.

\subsubsection{Data Reduction}
Data were reduced within IRAF\footnote{Image Reduction and Analysis
Facility (IRAF), a software system distributed by the National Optical
Astronomy Observatories (NOAO)}, using the standard {\tt DIMSUM} \citep{stan}
near-IR reduction package, including dark subtraction, flat fielding, sky-substraction, bad pixel masking, bright object masking, registration and summing.

VLT images were reduced with ISAAC pipeline recipes 5.4.2\footnote{Available at ftp://ftp.eso.org/pub/dfs/pipelines/isaac/} using the command line utility {\tt EsoRex}, which includes standard routines, including dark correction, flat field calibration, bad pixel cleaning, image correlation and recombination.

\subsubsection{Astrometry}
The astrometric calibration of the images was performed using the WCStools 
package\footnote{Available at  ftp://cfa-ftp.harvard.edu/pub/gsc/WCSTools/} \citep{wcs},
using the Two Micron All Sky Survey (2MASS) Catalogue of Point Sources a reference catalog \citep{cutri} and the USNO-A2.0 Catalogue \citep{monet} 
for fields with a few number of 2MASS sources. We estimate the uncertainty in the relative astrometry to be $\sim$0\farcs4-0\farcs6 .

\subsubsection{Photometry}
The photometry for the CTIO observations was calibrated using 2MASS point sources detected in these fields.
For VLT runs, magnitudes were calibrated using NICMOS near-infrared standards \citep{persson}.  
All standard star magnitudes were obtained using SExtractor \citep{bertin} 
with an aperture diameter of $6\arcsec$,  
which proved to be adequate by monitoring the growth curve of all the measured stars.
For the observations made with OSIRIS and CIRIM at CTIO the $K$--band zero point was 22.64$\pm$0.06 and 22.01$\pm$0.02 
(for 1 count/second, integrated over the source), respectively.
For the ISAAC/VLT observations we derive the zero-point based on observations of the near-IR photometric standard star
S301-D taken from the list of \citet{persson}. The zero-point obtained was 24.18$\pm$0.01.
The zero-points do not include the airmass term, because the airmass dependence in $K$--band is small and we have observed all 
our objects with airmasses $<1.3$, with a mean of 1.06.

Because of the large uncertainties in the photometry, we did not correct the $K$--band magnitudes for Galactic extinction. 
We have used the NED Database\footnote{http://nedwww.ipac.caltech.edu/ - the NASA-IPAC Extragalactic Database},  
which are based on the $E(B-V)$ values from the extinction maps of \citet{sch}. We found that most of the sources have values 
from 0.01 to 0.1, which are negligible compared to magnitude errors.

\section{Source detection and Near-IR identification}

We have identified the near-IR counterparts to the radio sources by overlaying the radio contour maps on to the $K$--band images using the \aips\ task {\tt KNTR}.

We measured the magnitudes of these identifications using SExtractor. The parameters were set such that, to be detected, an object must have a flux in excess of 1-1.5 times the local background noise level over at least $N$ connected pixels, according to the seeing conditions and image qualities.
In order to improve the detections of faint sources, we smoothed the $K$--band images using a circular Gaussian of FWHM=1-2 pixels.
In some cases, we find that it is possible that the identification does not fall at the midpoint of the radio lobes, for radio sources with double or multiple components.  

We measured magnitudes for all radio source identifications using SExtractor and circular apertures with diameters of 2\farcs0, 4\farcs0 and 8\farcs0. We also obtained {\tt MAG\_BEST} magnitudes in order to facilitate comparison with the literature \citep{bornan04,bornan06}. This is usually equal to {\tt MAG\_AUTO}, but if the contribution of other source exceeds 10\%, it is {\tt MAG\_ISOCOR}.  The uncertainties quoted represent our best estimates, which includes the zero-point uncertainty (See Table 2).

Table 2 lists our USS sample, the columns are: IAU J2000 designation, spectral index obtained between 352 MHz and 1.4 GHz, largest angular size, as measured from radio maps taken with the VLA, aperture magnitudes in  2\farcs0, 4\farcs0 and 8\farcs0, SExtractor {\tt MAG\_BEST} magnitudes, J2000 position coordinates of the radio source and the $K$--band identification counterpart, the half-light radius and {\tt CLASS\_STAR} values of the $K$--band identifications.
SExtractor {\tt CLASS\_STAR} parameter provides an indication of the likelihood of an object being a galaxy or a star. In the ideal cases a galaxy has {\tt CLASS\_STAR}$=0.0$ and a point source has {\tt CLASS\_STAR}$=1.0$. We find that all the USS sample analyzed in this work are consistent with extended sources (galaxies) rather than point sources (stars or quasars).

We also considered studying the environments of these USS radio galaxy candidates. The environments of twenty USS sources observed with CTIO/OSIRIS were analyzed in \citet{bornan04}. In those images, only 7 radio galaxies were identified, finding a statistically significant signal of clustering. In \citet{bornan06}, a similar analysis was performed on a wider field and with deeper data, finding a stronger result. We therefore do not analyze the environments of the USS sources in the relatively shallow $K-$band images from this paper.

\section{Optical Spectroscopy}

During the past 6 years, we have obtained optical spectroscopy of 14 USS
sources. The targets were mostly selected from 4 different USS samples:
the 3 sub-samples of De Breuck et al. (2000) and the WISH--NVSS sample
described in this paper. From these samples, we mostly selected the final
targets based on the RA range when telescope time was available.

\subsection{VLT}

Optical spectroscopy of a subsample of USS sources were carried with FORS2 installed on the VLT Unit Telescope 4 Yepun between 2000 October and 2001 May.
We used a similar data reduction strategy to \citet{DB06}.
We removed cosmic rays using the IRAF task {\tt szap}, after bias and flatfield correction.
We used the IRAF task {\tt background} in order to remove sky variations and we extracted the spectra using a width appropriate to contain all flux in the extended emission lines. 
Finally we calibrated the one dimensional spectra in wavelength and flux.
We used the procedures described in \citet{roet97} and in \citet{DB06} to determine the central wavelength, total line flux, deconvolved widths, and rest-frame equivalent widths (with their related errors) of all the emission and absorption lines detected in our spectra. For each object, we quote a single redshift, which is an average of all the lines, weighted by the uncertainties due to the line fitting and wavelength calibration.
In Table 4 we show the measured parameters of the emission/absorption lines in the spectra. 

\subsection{Keck}

Longslit observations were made on UT 2004 January 19 - 20, using the Low
Resolution Imaging Spectrometer \citep[LRIS;][]{lris} on Keck I, with a slit
width of 1\farcs5, under photometric conditions and 0\farcs8 seeing. The
spectroscopic setup is described in Table 3. Data reduction was performed in
IRAF; data were bias-subtracted, response-corrected, cosmic-ray-zapped using
Pieter van Dokkum's LACosmic routine
\footnote{{http://www.astro.yale.edu/dokkum/lacosmic/}}, sky-subtracted,
and spectra extracted in a 1\farcs5 aperture. The resultant 1-D spectra were
wavelength and flux calibrated using appropriate arcs and standard stars.

\subsection{WHT}

On UT 2006 December 14 and 15, we obtained 3 spectra using the ISIS double beam 
spectrograph \citep{car94} at the 4.2\,m William Herschel telescope (WHT) at 
Observatorio Roque de los Muchachos on the Canary island of La Palma. 
Conditions were photometric with 0\farcs5 and 1\farcs5 seeing during the first 
and second night, respectively. We used a dichroic splitting the light at 
5300~\AA\ and a 1\farcs5 wide slit. The grisms used were the R300B in the blue 
arm and the R316R in the red arm. None of the 3 objects were detected in the 
blue arm. In the red arm, only WN~J0912$-$1655 was clearly detected in 5400\,s 
of observing time. WN~J0604-2015 and WN~J0610+6611 were not detected in 5400\,s 
and 7200\,s, respectively. We extracted the red spectrum of WN~J0912$-$1655 with 
an aperture width of 2\farcs4.

\subsection{Results}
We have determined redshifts of 12 out of 14 targets observed. The two targets without redshifts were undetected in 1.5 to 2 hours on a 4\,m telescope, and need long integration on 8--10\,m class telescopes. Apart from TN~J2009$-$3040, which is a $z$=3.158 quasar, all 11 targets are radio galaxies with redshifts from $z$=0.550 to $z$=3.837. Five of the radio galaxies are at $z$$>$2, including three at $z$$>$3. Table~4 lists the emission line parameters of all targets.
The 10 new redshifts from the De Breuck et al. (2000) USS sample raises
the number of spectroscopic observations from that sample from 46 to 56,
and the number of known redshifts from 34 to 44.

\section{discussion}

\subsection{Correlations}
In this work we have excluded large radio sources because those are most likely foreground objects and not at high redshift. 
In Figure 1 we present the correlation between largest angular size (LAS) and the $K-$magnitudes obtained within an aperture of 4$\arcsec$. As can be seen most of the sources have small radio sizes with LAS $<$ 8$\arcsec$.
The vertical dashed line represent the expected $K-$band magnitude of a $z\sim3$ source, indicating that high redshift sources have compact radio morphologies.
Several authors have applied the small angular-size criterion in order to find high redshift radio galaxies \citep{blundell,jarvis04,cruz}. There are a few cases of radio galaxies at high redshifts with large angular sizes (4C 23.56 at z=2.483 has LAS=53$\arcsec$). However, \citet{pedani} found that the introduction of a strong angular size bias such as LAS$<15\arcsec$, increases the efficiency in selecting high redshift radio galaxies by a factor of two. 

Some radio galaxies have very faint or no $K$--band counterpart presenting large deviation in the Hubble $K-z$ diagram \citep{DB06}, and 20\% of high-redshift radio galaxies fail to show optical emission lines in spectroscopy observations using the Keck Telescope \citep{DB01,reulandb}. This could be an indication that some high redshift radio galaxies contain great amounts of dust.
\citet{willott02} found an anti-correlation between dust emission (measured from 850 $\mu$m flux) and the projected linear size, though \citet{reuland04} did not confirm this using a larger sample of radio galaxies. 
It is likely that the compact USS sample analyzed here may also contain a mix of galaxies either at very high redshift, or heavily obscured by dust.

\begin{figure}
\includegraphics[width=80mm]{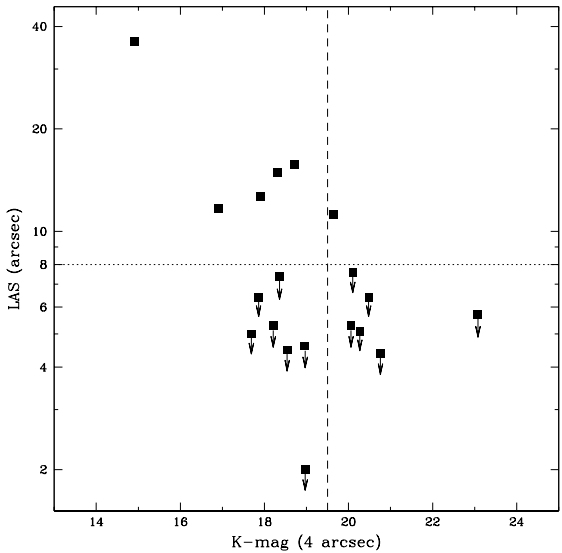}
\label{fig1}
\caption{Radio largest angular size vs $K-$magnitudes (4$\arcsec$ aperture).
Dashed line represent the expected $K-$band magnitude of a $z\sim3$ source. Most of the sources have small radio sizes as indicated by the horizontal dotted line.}
\end{figure}

\begin{figure}
\includegraphics[width=80mm]{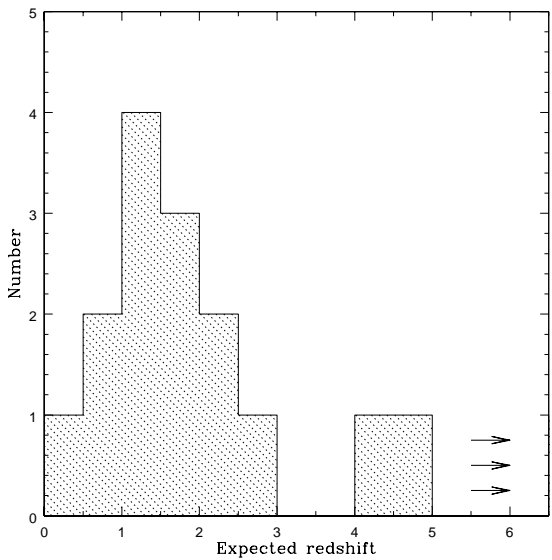}
\label{fig2}
\caption{Distribution of expected redshift of USS sources, estimated from the Hubble $K-z$ diagram using 64 kpc metric apertures.}
\end{figure}

\subsection{Expected redshift distribution}
We have estimated the redshift of the USS sample using linear regression and the Hubble $K-z$ diagram of \citet{DB02}. Using aperture photometry in a 64\,kpc metric aperture, \citet{wish} found $K=$4.633$\times$log$_{10}(z)-17.266$.
As we expect most radio galaxies to have $1<z<4$, we have
assumed $K_{64kpc}=K(8\arcsec)$, as 8$\arcsec$ corresponds to $\sim$64\,kpc at $z$=1, which only a small dependence on redshift $z>1$ in our adopted Cosmology.
Figure 2 shows the expected redshift distribution for our USS sample. The median predicted redshift is $\overline{z}_{exp}$=2.13, which is higher than the predicted redshift obtained by \citet{sumss} for the SUMSS--NVSS sample ($\overline{z}_{sumss}=1.75$) and the expected redshift obtained in the 6C$^{*}$ \citep[$z\sim1.9$;][]{jarvis01} and 6C$^{**}$ surveys \citep[$z\sim1.7$;][]{cruz07}.

In Figure 4 we show the $K-$band magnitude distribution of USS sources, measured in a 8\arcsec diameter aperture. We compare our results with those obtained in the SUMSS--NVSS \citep{sumss}, 6C$^{**}$ \citep{cruz07} catalogues and data selected at 74 MHz taken from \citet{jarvis04}.  
The mean $K-$band magnitudes in an 8$\arcsec$ diameter aperture of the USS is $\overline{K}$=18.58, which is one magnitude fainter than those obtained by \citet{sumss}  in the SUMSS--NVSS survey ($\overline{K}$=17.57) and by \citet{cruz} in the the 6C$^{**}$ survey ($\overline{K}$=17.59).
We find a similar result using {\tt BEST} magnitudes. Our USS sample has $\overline{K}_{\tt BEST}$=18.74, and from a similar study \citet{bornan06} found $\overline{K}_{\tt BEST}$=17.97. 
The fainter $K$-band counterparts of the WISH--NVSS USS sources could be an indication that the host galaxies are located either at higher redshifts (i.e. a more efficient selection), or in a very dense, dusty medium. 

Alternatively, the WISH--NVSS USS radio galaxies are intrinsically less
massive and hence less luminous than the ones in previous samples. This
could be due to the fainter flux limits probed in this sample. However,
using {\it Spitzer} rest-frame near-IR phometry, Seymour et al. 2007 do
not find a strong dependence of host galaxy mass on 3\,GHz radio
luminisity, so we regard this explanation as unlikely given the small
difference in radio luminosity between the USS samples.

Alternatively, the WISH--NVSS USS sources may be less massive and hence
fainted in the observed $K-$band. This may be because WISH--NVSS
sources have fainter radio fluxes than most previous USS samples. Such a
radio power dependence in the $K-z$ diagram has been suggested by various
authors \citep{eales97,best98,jarvis01,DB02,willott02}. However,
using {\it rest-frame} near-IR imaging of a sample of 69 radio galaxies at
$z>1$ observed with {\it Spitzer}, Seymour et al. (2007) do
not find a strong dependence of host galaxy mass on 3\,GHz radio
luminisity, so we regard this explanation as unlikely given the small
difference in radio luminosity between the USS samples.

\begin{figure}
\includegraphics[width=80mm]{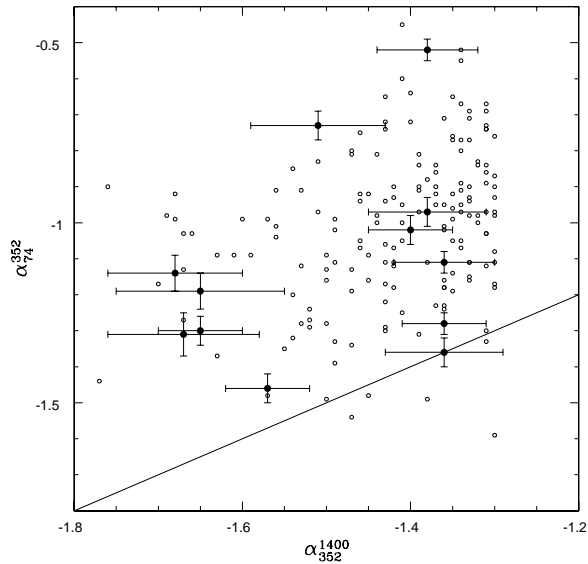}
\caption{Radio two-colour diagram for USS sources in our sample (filled circles). 
Open circles represent measurements obtained using the WENSS--NVSS sample at 325 MHz and 1.4\,GHz taken from \citet{DB00}. 
The line indicates the relation for USS sources whose spectra follow a single power law from 74 to 1400 MHz. }
\end{figure}

\begin{figure}
\includegraphics[width=80mm]{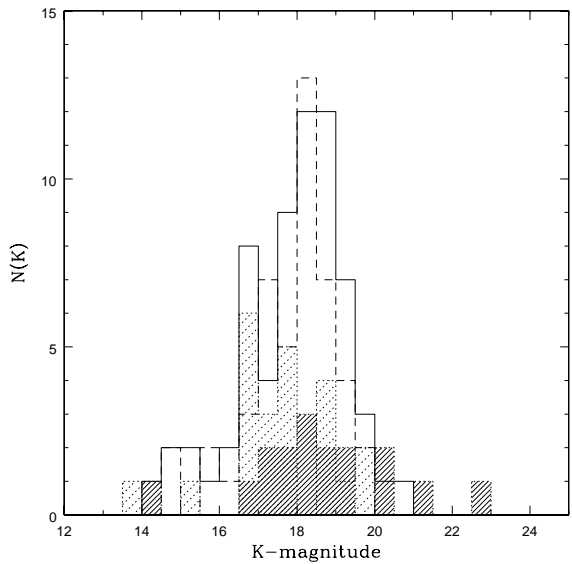}
\label{fig3}
\caption{$K-$band magnitude distribution of USS sources, measured in a 8\arcsec diameter aperture (shaded histogram). 
Solid and dashed lines represent the distribution of $K-$band magnitude, measured in a 8\arcsec diameter aperture for sources detected in the SUMSS--NVSS \citep{sumss} and 6C$^{**}$ \citep{cruz07} catalogues. Dotted histogram represent USS sources selected at 74 MHz taken from \citet{jarvis04}.  }
\end{figure}

\subsection{Radio two-colour diagram and spectral curvature}

We use the CATS database of the Special Astronomy Observatory \citep{verkho} to search for radio measurements in order to study the radio spectral energy distribution at low frequencies.
Figure 3 shows the radio two-colour diagram, which compares the spectral indices at 74-352 MHz and 352-1400 MHz for 12 USS sources of the total sample (Filled circles). Fluxes at 74 MHz were obtained from the VLA Low-frequency Sky Survey \citep{cohen}.
The line indicates the relation for USS sources whose spectra follow a single power law from 74 to 1400 MHz.
Open circles represent measurements obtained using the WENSS--NVSS sample at 325 MHz and 1.4\,GHz taken from \citet{DB00}.
We find a general agreement with those values. 
Only a single source in our sample is consistent with a straight spectrum, and it is clear that most of the USS sources have substantially flatter spectral indices between 74 and 352 MHz than between 352 and 1400 MHz. 
The median spectral index obtained at 74-352 MHz is $\overline{\alpha}_{74}^{352}=-$1.1, while the median at high frequencies is $\overline{\alpha}_{352}^{1400}=-$1.5. The spectral flattening is in the range 0$<$$\Delta(\alpha)$$<$0.8. Unfortunately, we do not have redshifts for any of the 12 sources with 74\,MHz data. However, WN~J2007-1316 at $z$=3.837 is detected in the Mauritius Southern Sky Survey \citep{pandey} with $S_{\rm 151MHz}$=3.58$\pm$0.57\,Jy. This implies $\alpha_{151}^{352}$=$-$1.60$\pm$0.19, compared to $\alpha_{352}^{1400}$=$-$1.52$\pm$0.04. Although the uncertainty in the spectral index is larger than for the 74-352 MHz data, this result suggests that the highest redshift radio galaxy confirmed in our sample does retain a straight radio spectrum through the lowest frequencies observed.
As noted by \citet{klamer} $\sim$90\% of USS sources selected from the SUMSS and NVSS catalogues show a straight radio spectra characterised by a single power law beyond $\sim$1 GHz.
At frequencies below 100 MHz spectral curvature is much more common than at higher frequencies due principally to synchrotron self-absorption. This suggests that very low frequency selected USS samples will be likely to be more efficient to find  high redshift galaxies.

\section{Conclusions}

We present $K$--band observations obtained with CTIO at Cerro Tololo and with VLT at Cerro Paranal, high-resolution VLA radio maps for a sample of 28 Ultra Steep Spectrum Radio Sources selected from the WISH catalog. We also present  optical spectroscopy from VLT/FORS2, Keck/LRIS and WHT/ISIS for 12 USS sources, mainly from the \citet{DB00} samples.

We find that most of our USS sources have small radio sizes and fainter magnitudes.
The mean $K-$band counterpart magnitude measured in  8$\arcsec$ diameter aperture is $\overline{K}$=18.58, which is one magnitude fainter than those obtained on the literature. 
The expected redshift distribution estimated using the Hubble $K-z$ diagram has a mean of $\overline{z}_{exp}$=2.13, which is higher than the predicted redshift obtained for the SUMSS--NVSS sample and the expected redshift obtained in the 6C$^{*}$ and 6C$^{**}$ surveys. 

From our spectroscopy, we identify one $z$=3.158 quasar, and 11 new radio galaxies, 3 of which are at $z$$>$3.

From a radio colour-colour diagram of a subsample of 12 sources, we find that all but one radio spectra flatten significantly below 352\,MHz. We also find that the highest redshift source from this paper (at $z$=3.84) does not show evidence for spectral flattening down to 151\,MHz.
Because most low redshift radio galaxies have a flatter spectra below 352 MHz, selecting USS sources at low frequencies is likely to be more efficient selection technique to find high redshift galaxies.
The forthcoming low frequency observatories such as LOFAR, LWA or SKA will be enable to find large samples of low-frequency selected USS sources \citep[e.g.][]{roet06} and will therefore open new perspectives on the nature and evolution of high redshift radio galaxies.

\begin{figure*}
\begin{tabular}{ccc}
WN~J0037$-$1904 $\sigma$=0.3 mJy & WN~J0141$-$1406 $\sigma$=0.1 mJy & WN~J0224$-$1701 $\sigma$=0.2 mJy  \\
\includegraphics[width=53mm,angle=270]{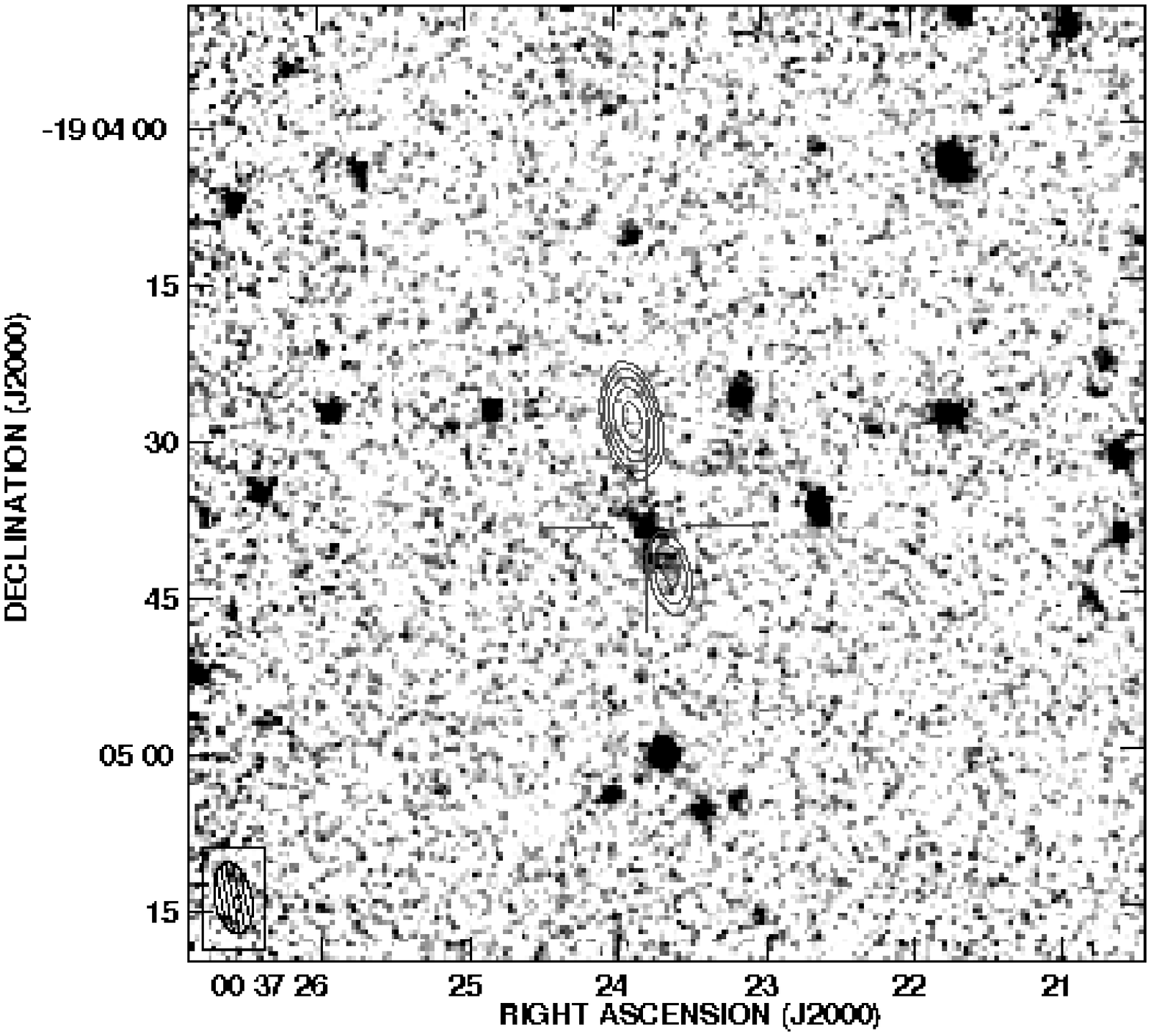}&
\includegraphics[width=53mm,angle=270]{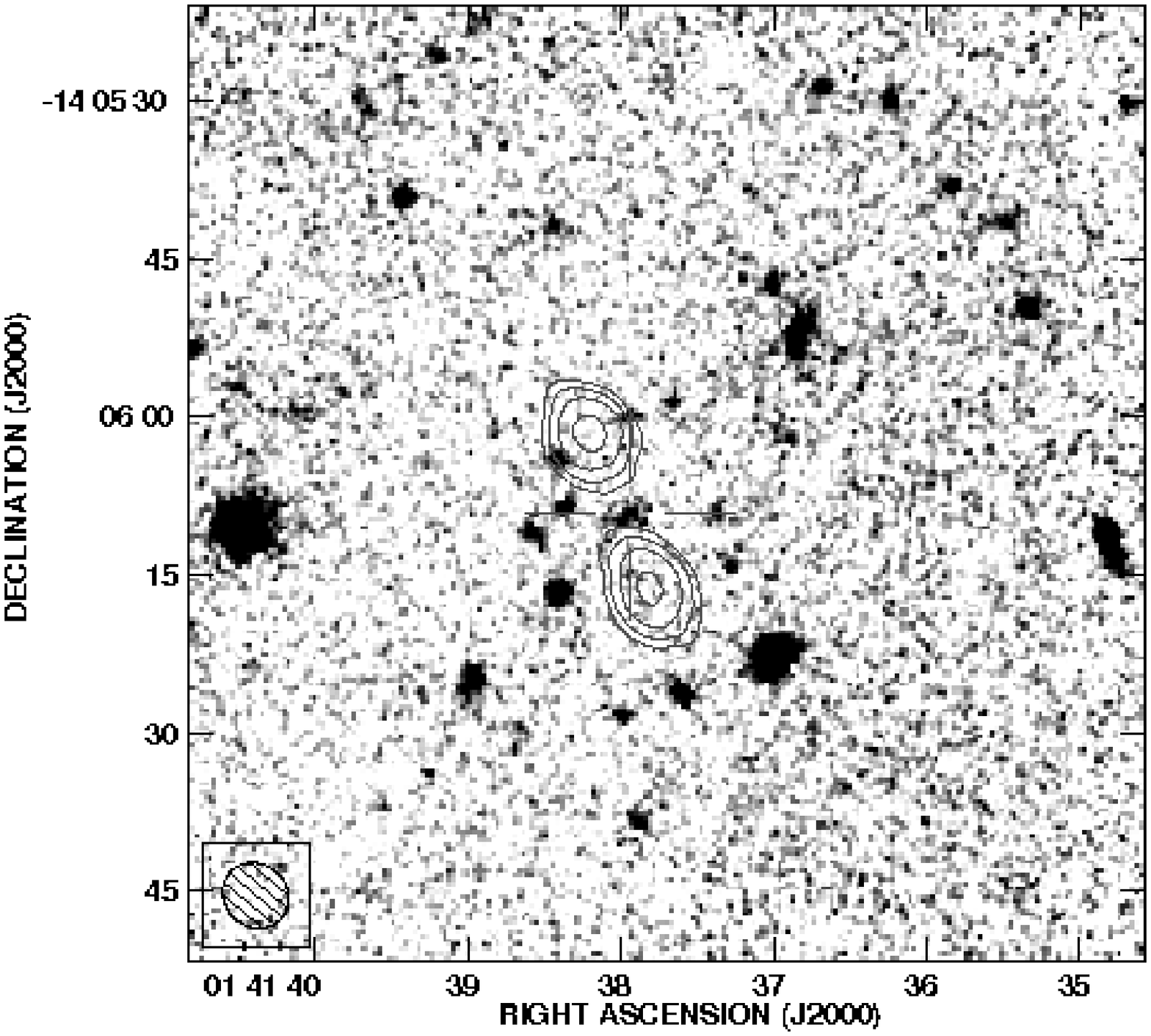}&
\includegraphics[width=53mm,angle=270]{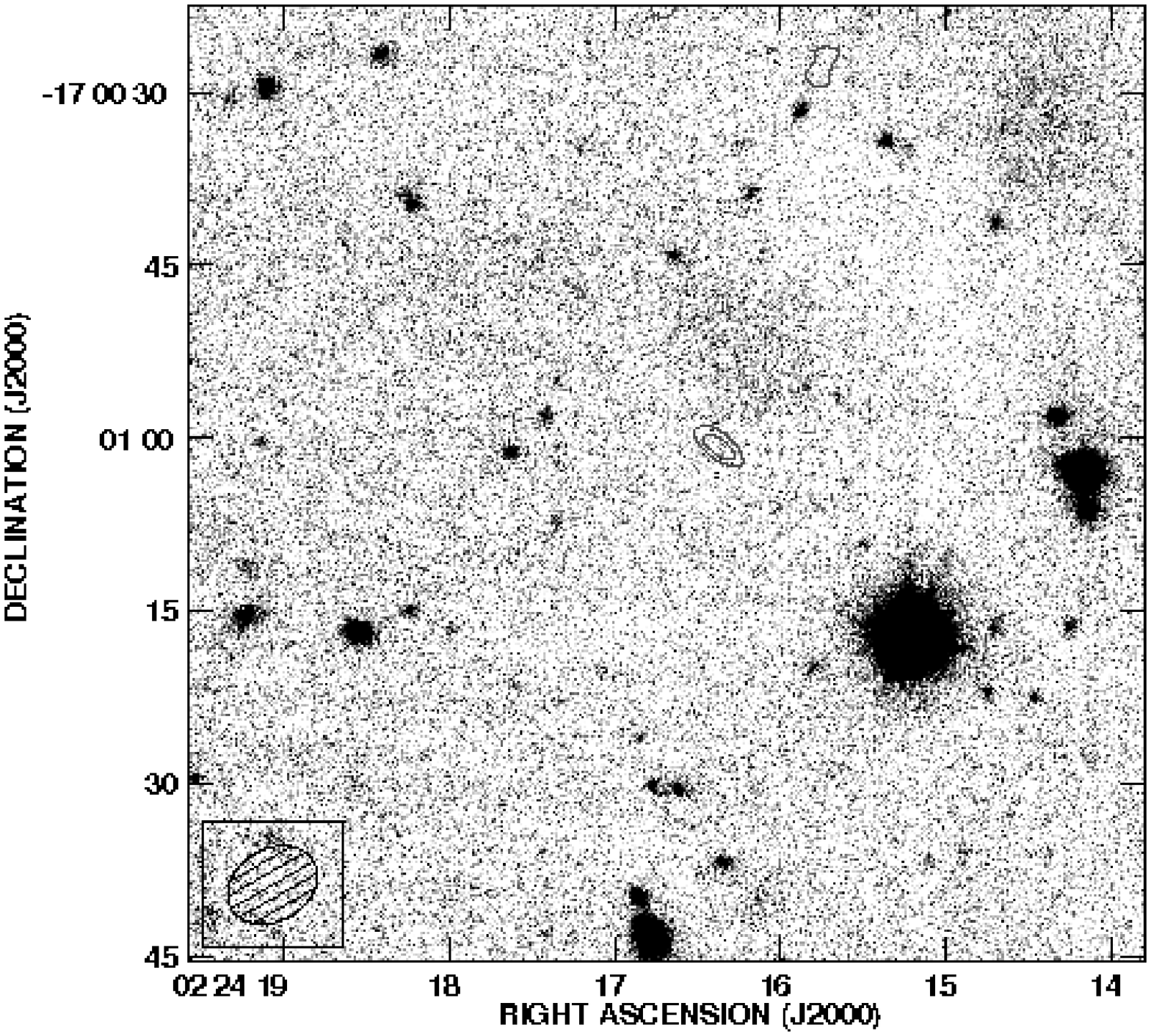}\\
WN~J0246$-$1649 $\sigma$=0.2 mJy & WN~J0510$-$1838 $\sigma$=4.8 mJy & WN~J0604$-$2015 $\sigma$=0.1 mJy \\
\includegraphics[width=53mm,angle=270]{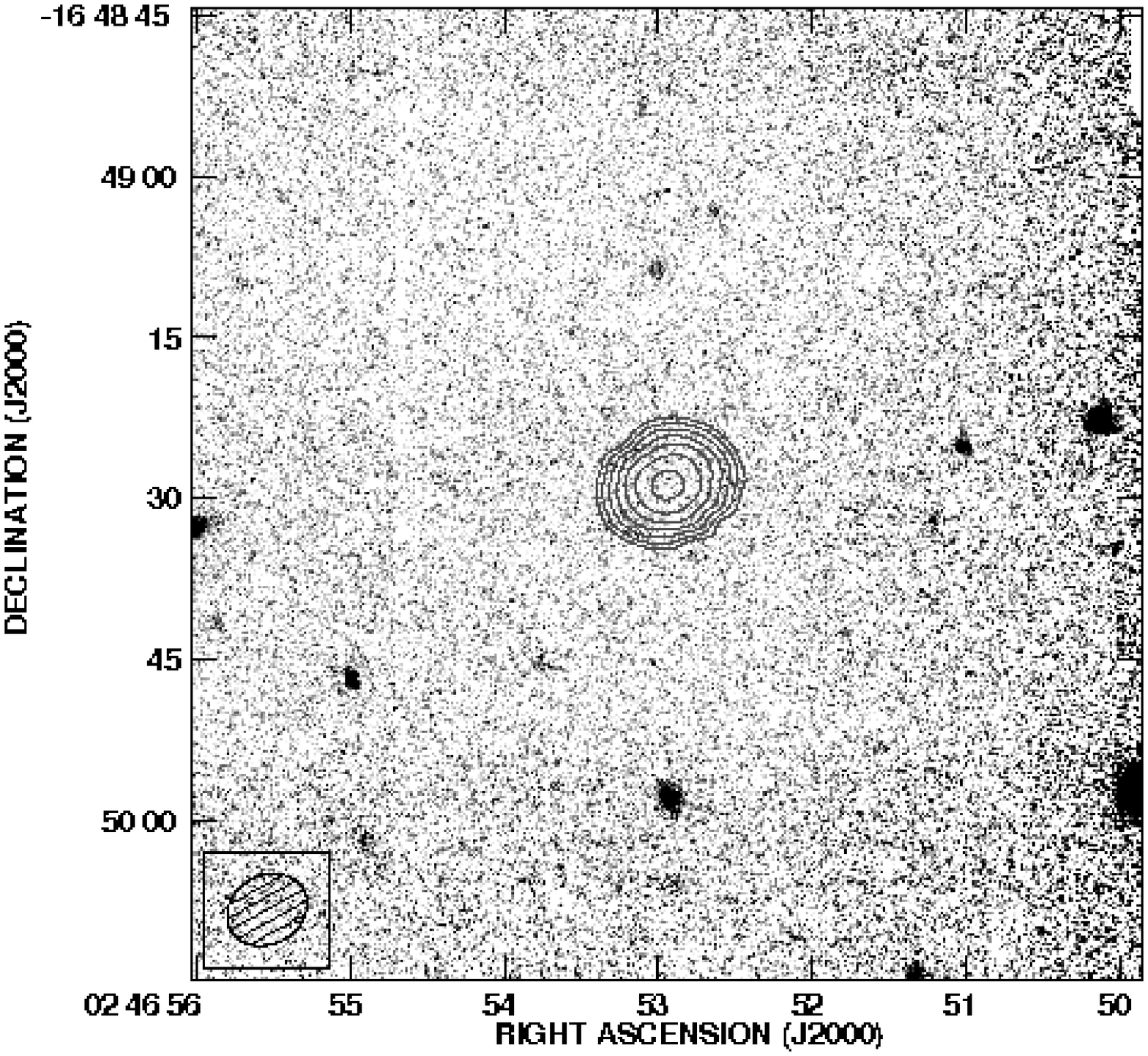}&
\includegraphics[width=53mm,angle=270]{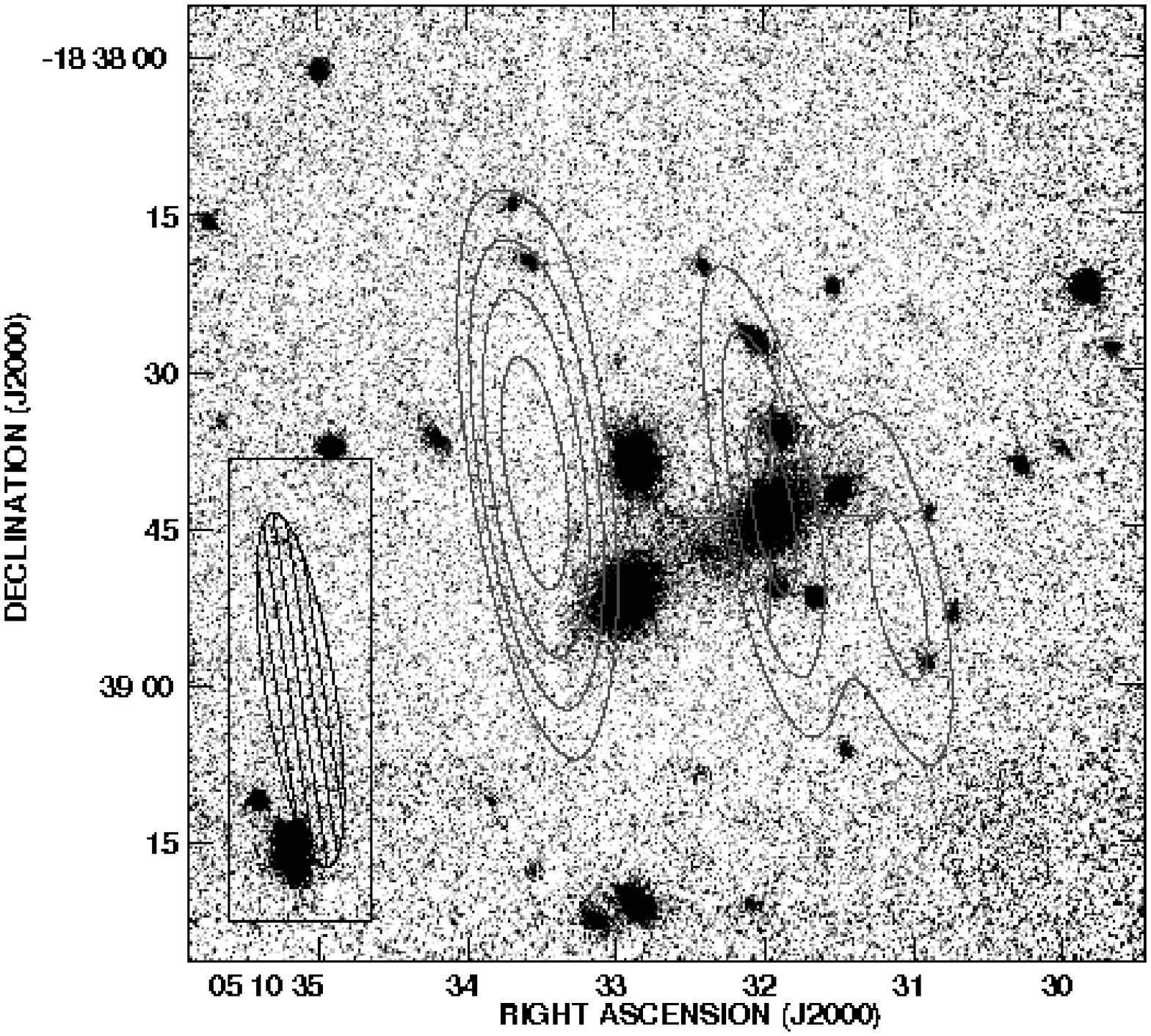}&
\includegraphics[width=53mm,angle=270]{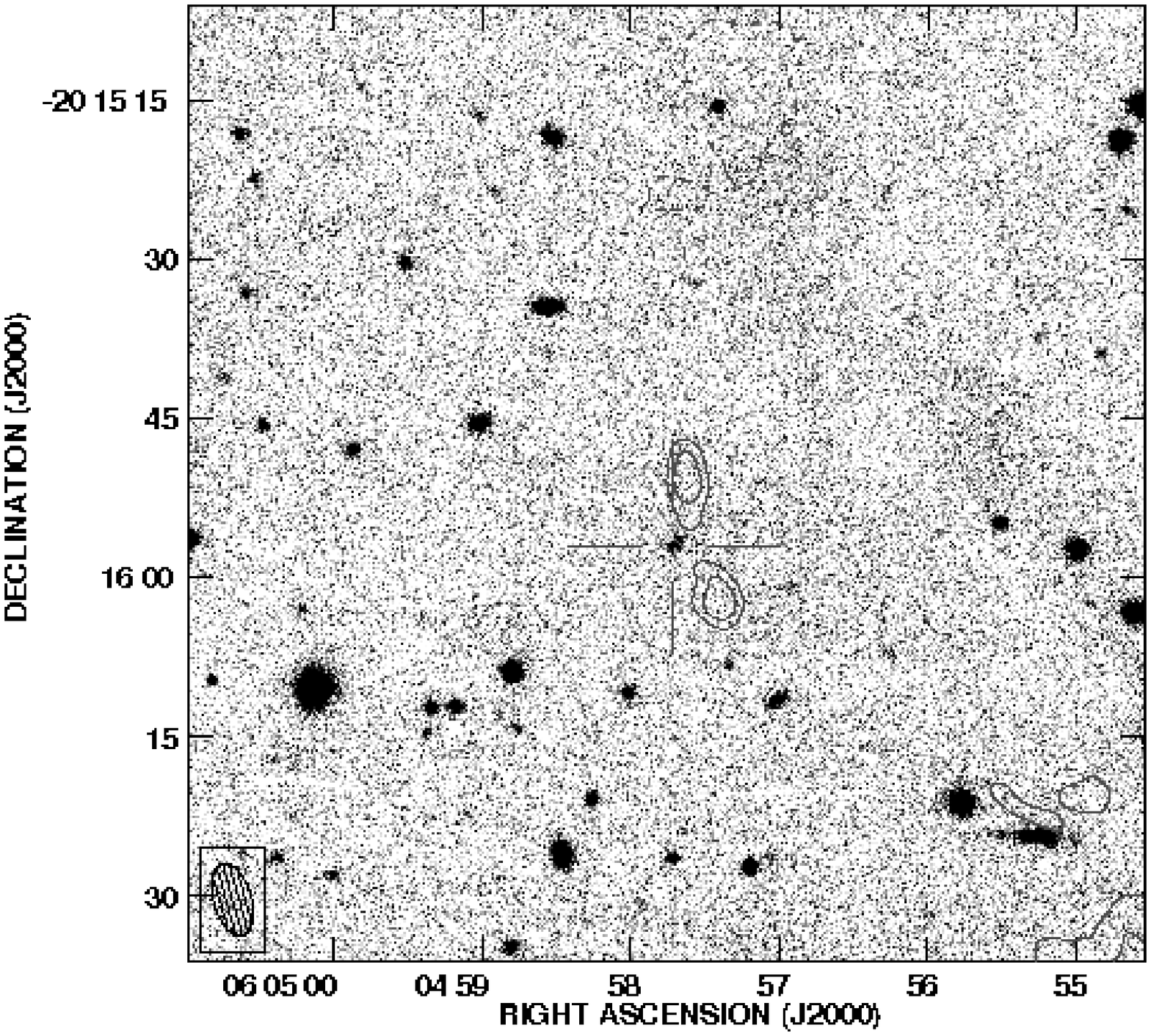}\\
WN~J1052$-$1812 $\sigma$=0.1 mJy $\sigma$=0.1 mJy &  WN~J1331$-$1947 $\sigma$=0.25 mJy & WN~J2007$-$1316 $\sigma$=0.1 mJy  \\
\includegraphics[width=53mm,angle=270]{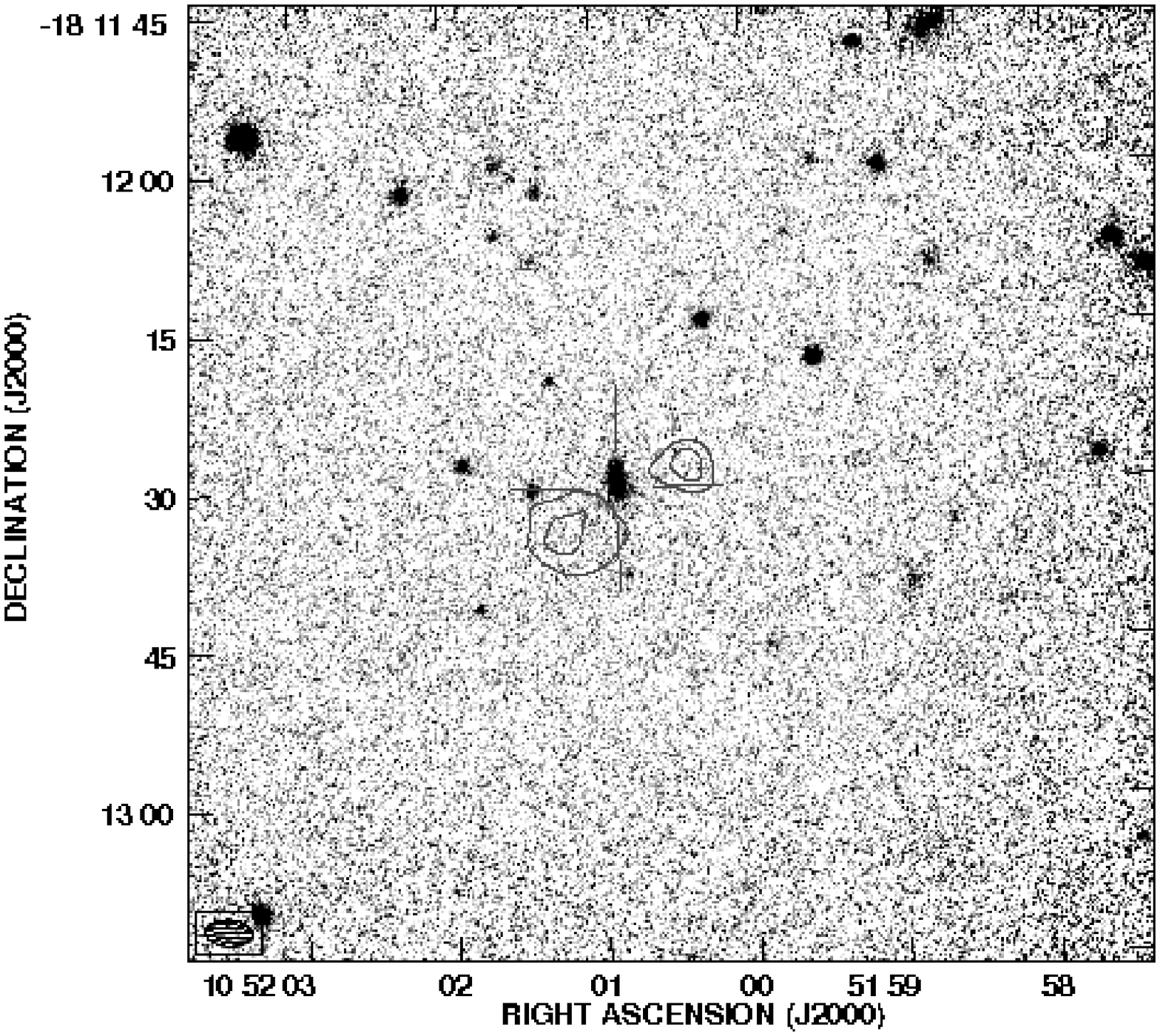}&
\includegraphics[width=53mm,angle=270]{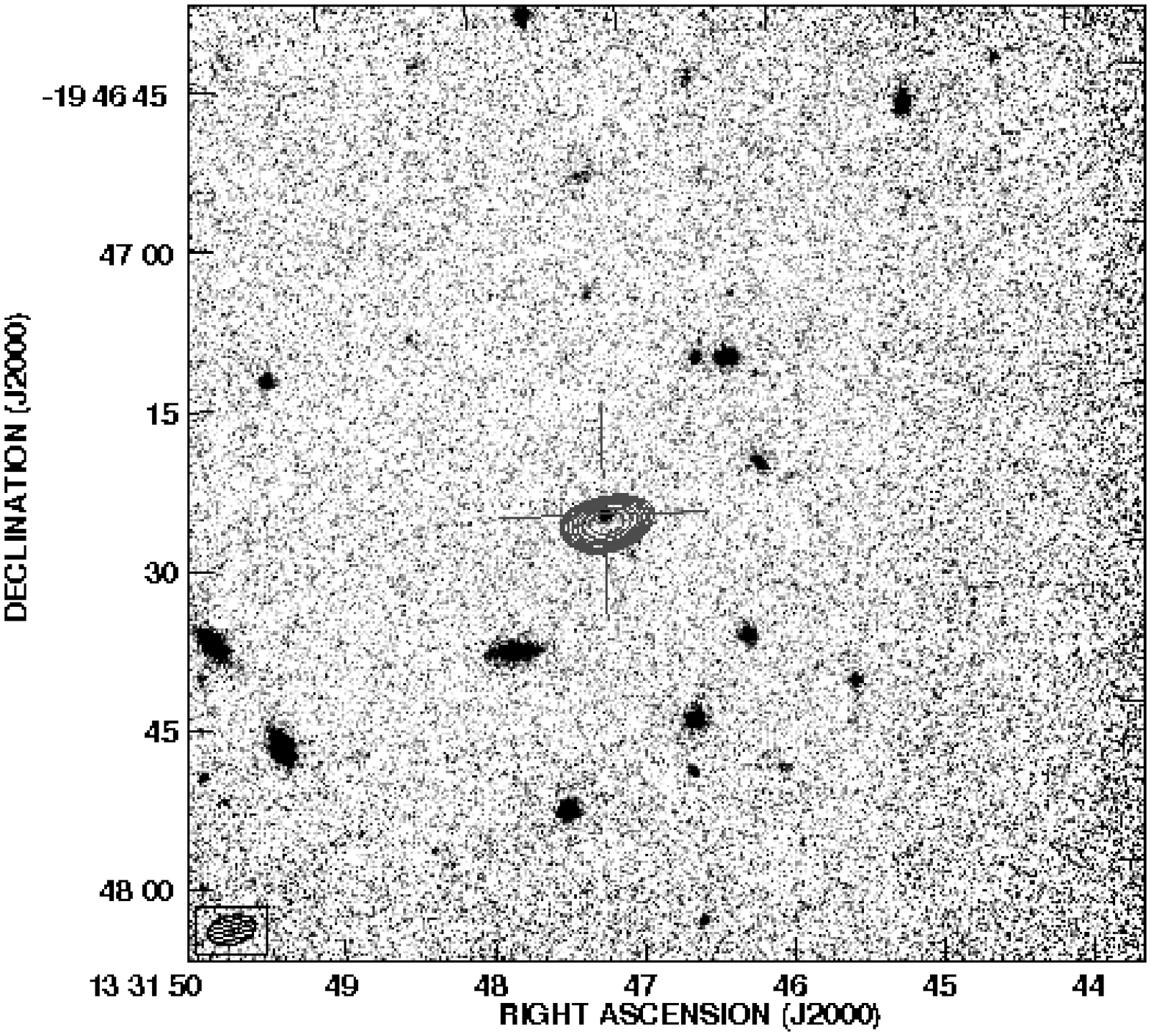}&
\includegraphics[width=53mm,angle=270]{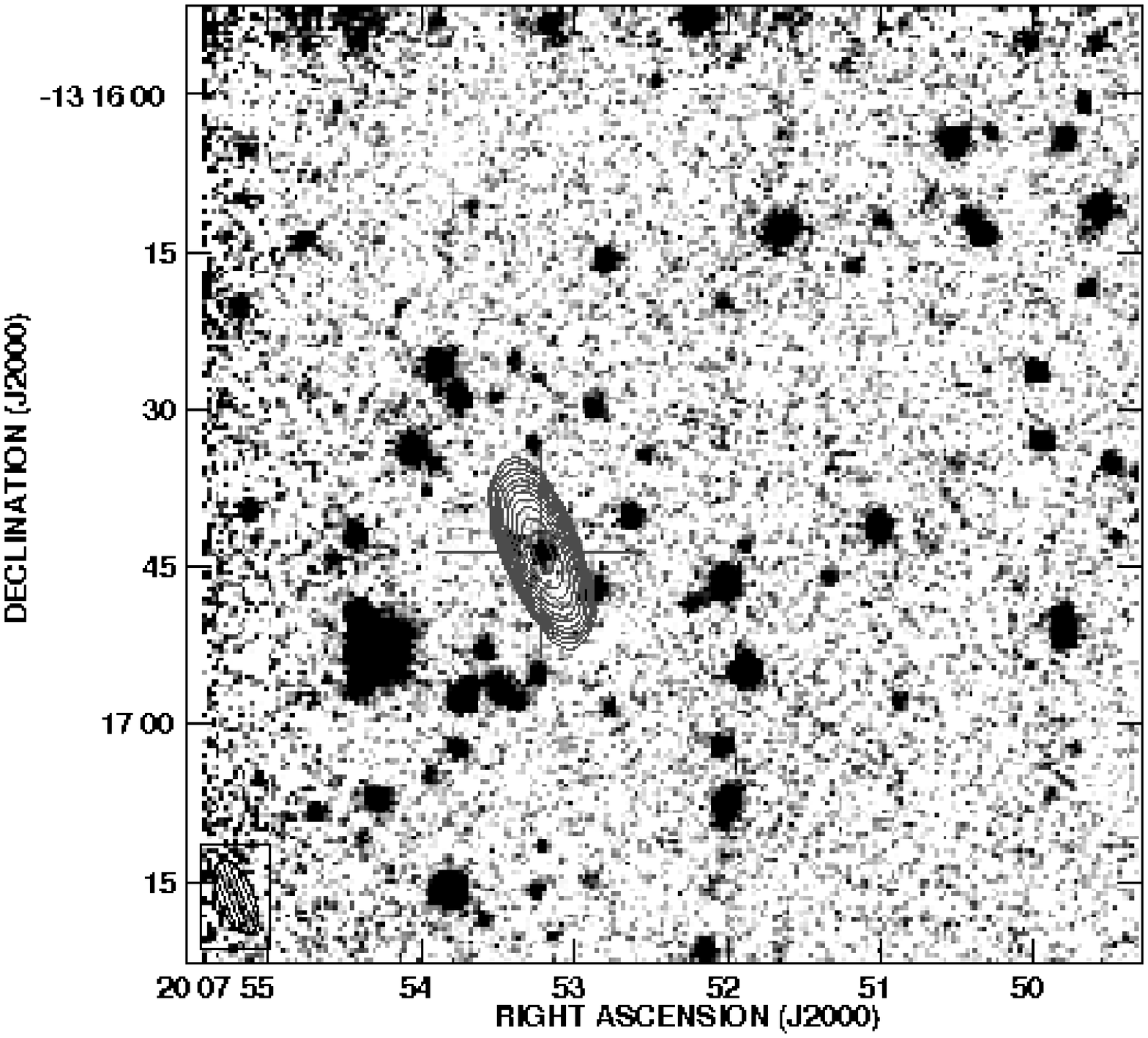}\\
\end{tabular}
\caption{Part of the $K$--band images of the WISH sample from observation of CTIO/OSIRIS, CTIO/CIRIM or VLT/ISAAC, with radio contours overlaid.
The contours represent the VLA radio emission at 1.4 GHz.
The contour scheme is a geometric progression in $\sqrt 2$, which implies
a factor two change in surface brightness every 2 contours. The first
contour level is at $3\sigma_{\rm rms}$,
where $\sigma_{\rm rms}$ is the rms noise measured around the
sources, which indicated above each plot. The half-power beamwidth 
 is indicated in the lower left corner of the
plots. The open cross indicates the $K-$band identification, as listed
in Table~2. See http://www.eso.org/$\sim$cbreuck/papers.html for a version with all 28 overlays}
\end{figure*}

\begin{figure*}
\begin{tabular}{ccc}
\centering
\includegraphics[width=9cm,angle=0]{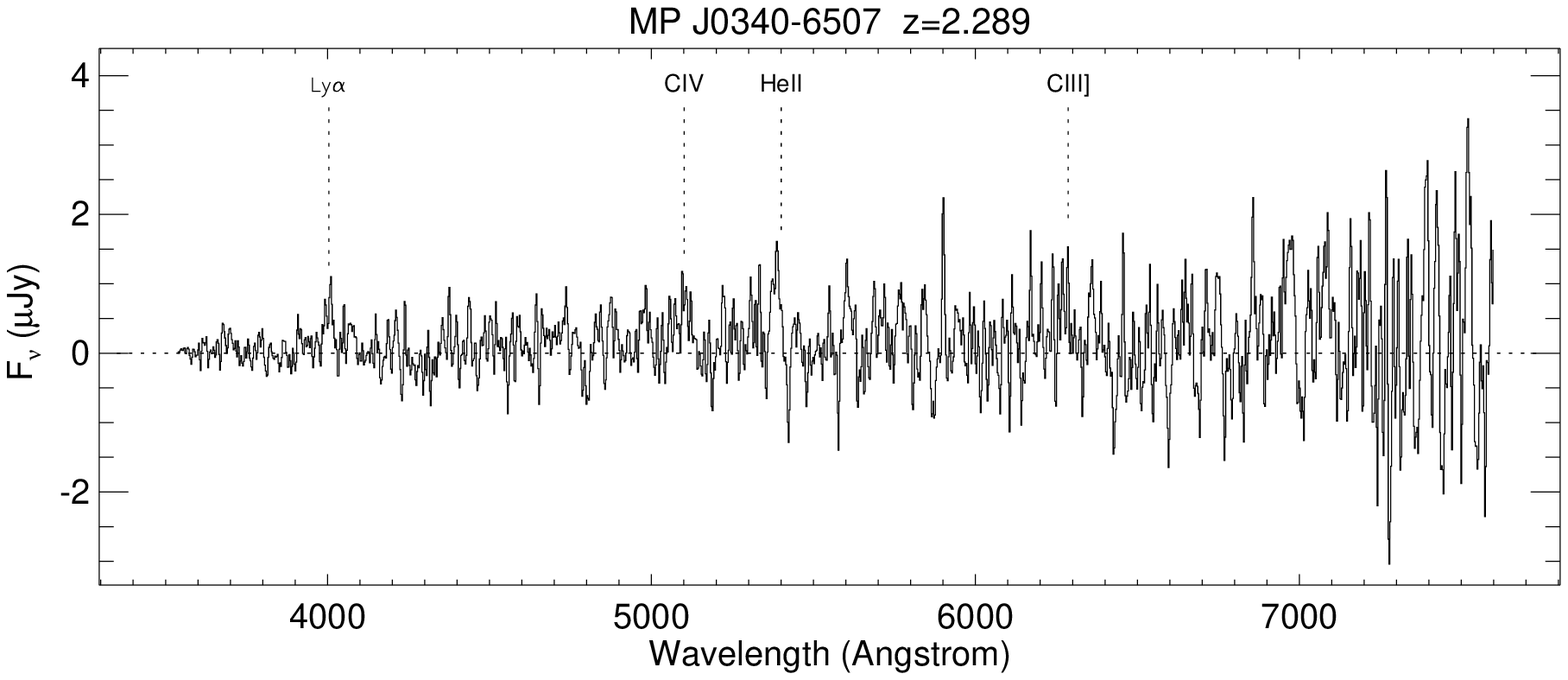}&
\includegraphics[width=9cm,angle=0]{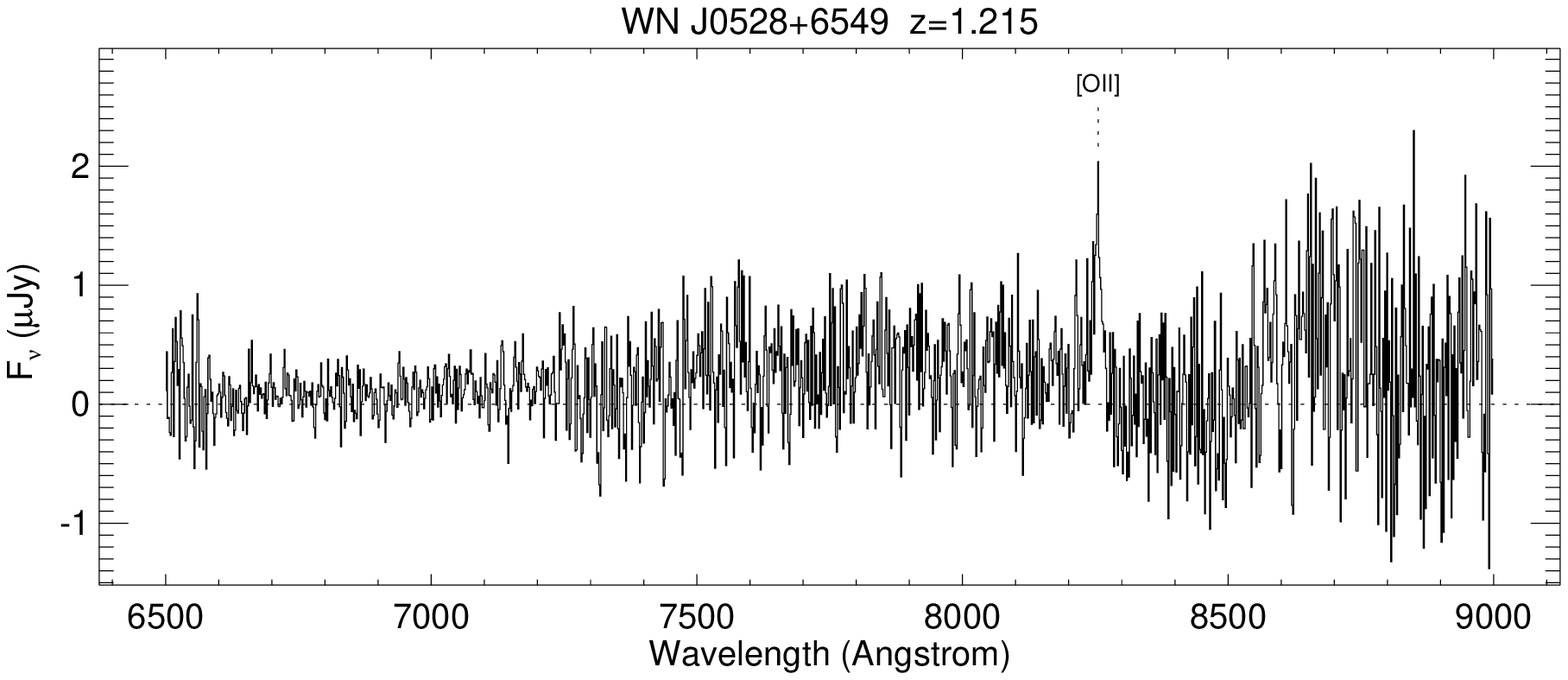}\\
\includegraphics[width=9cm,angle=0]{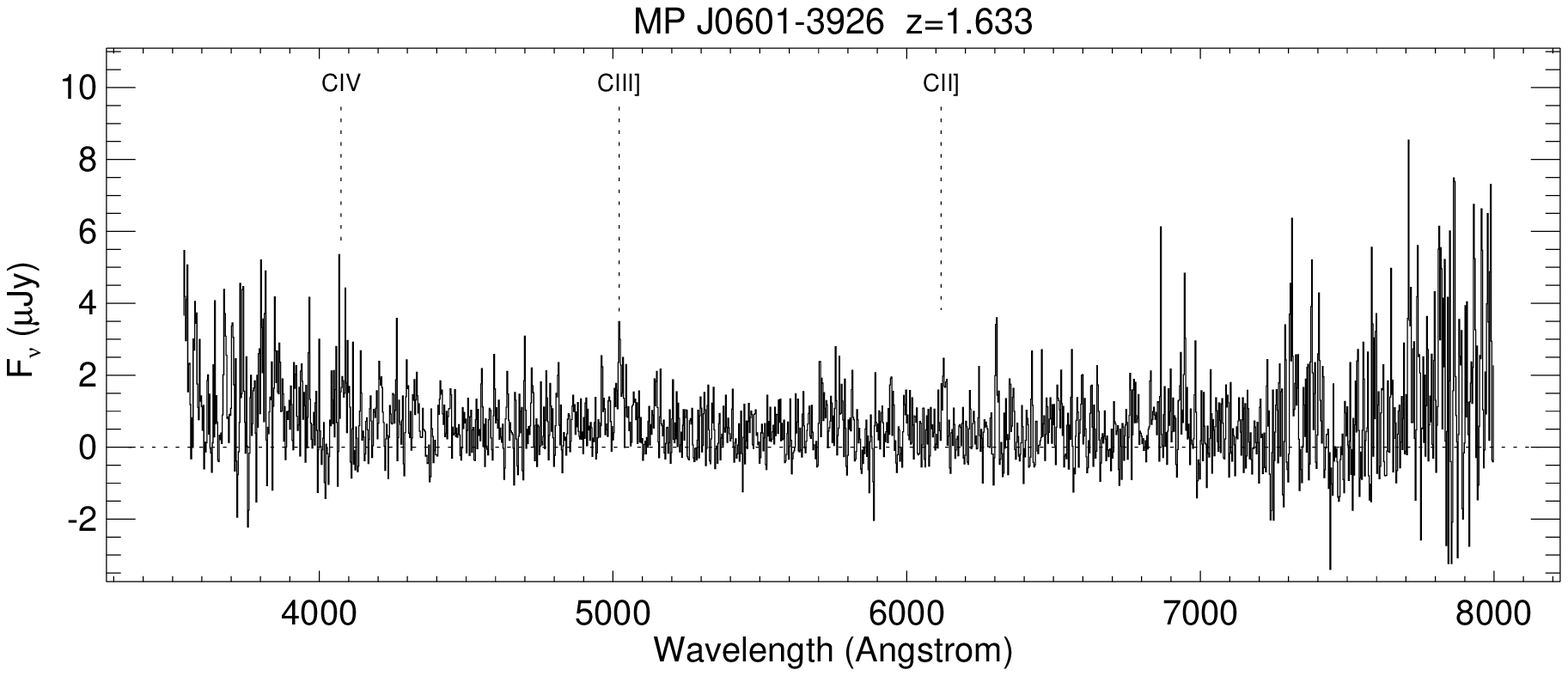}&
\includegraphics[width=9cm,angle=0]{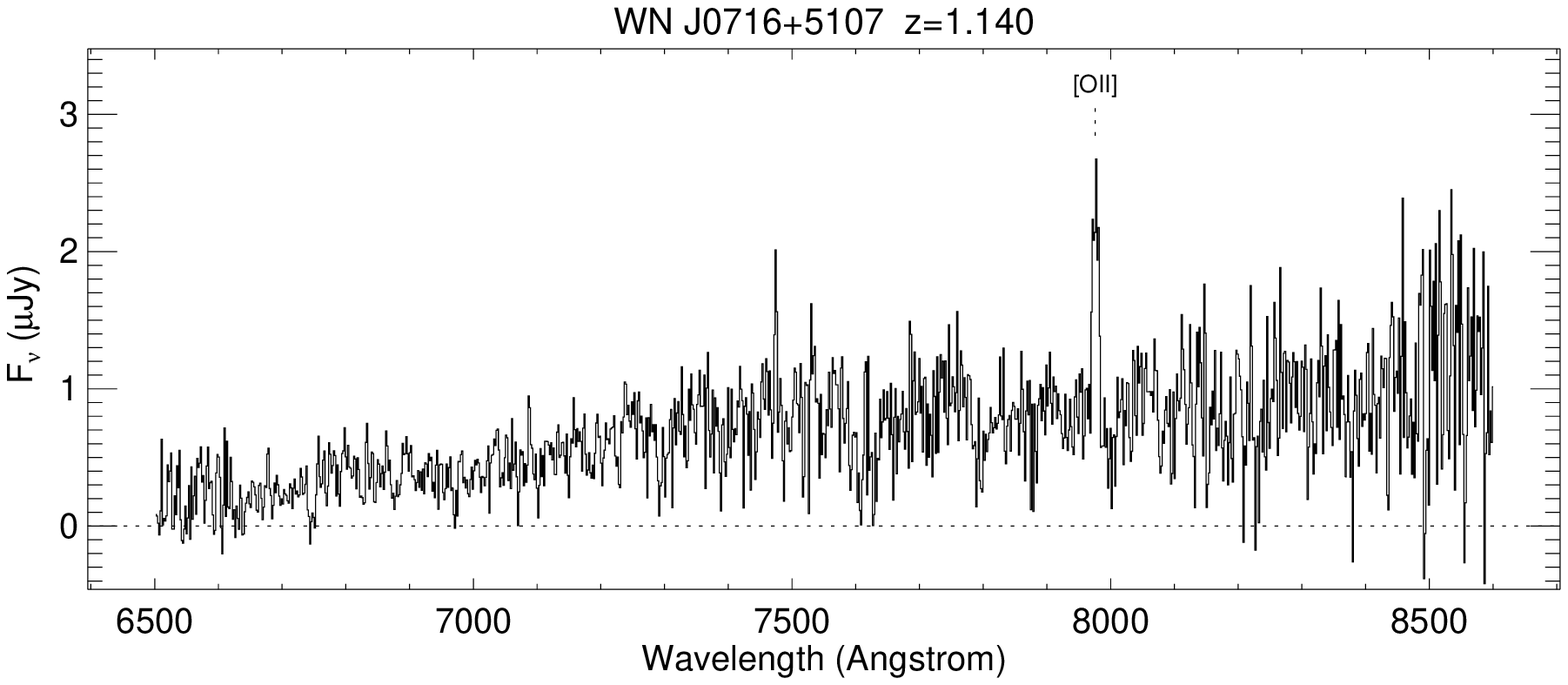}\\
\includegraphics[width=9cm,angle=0]{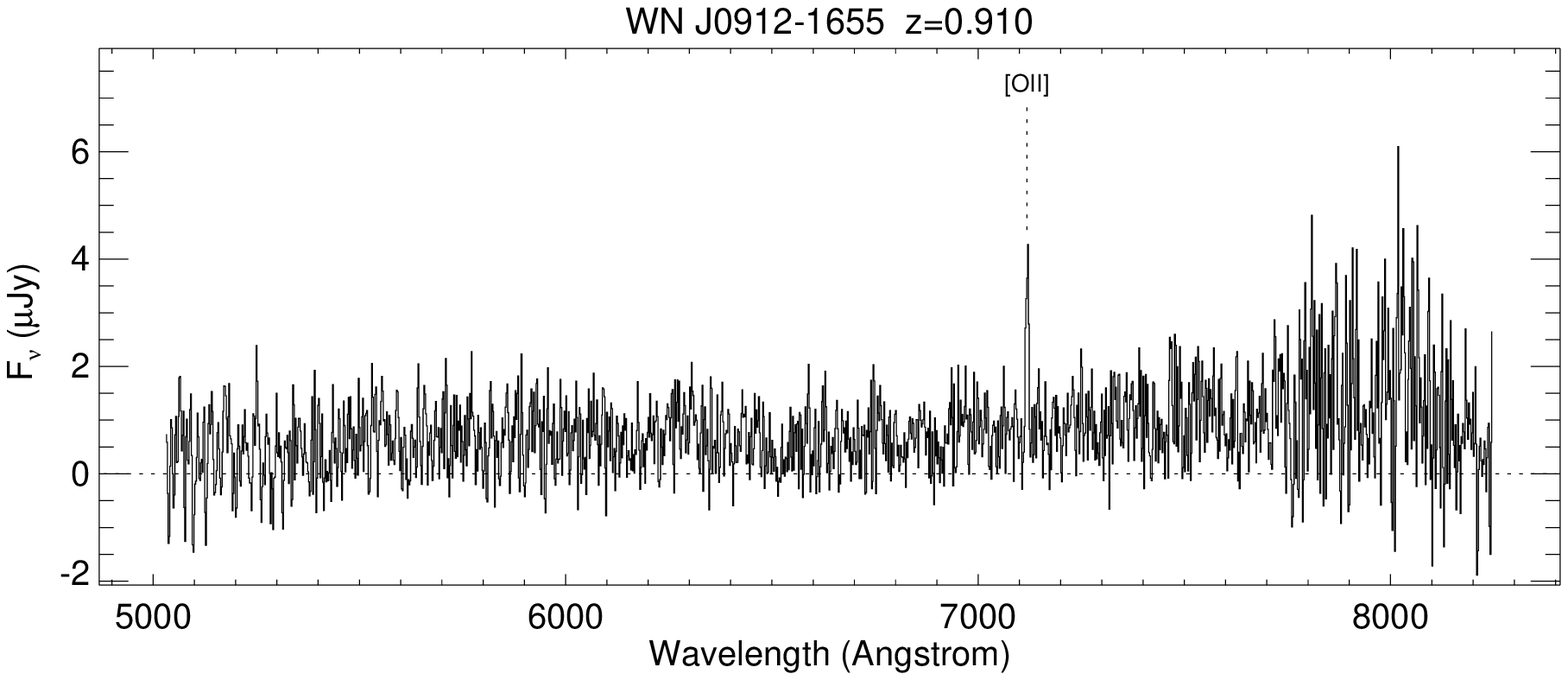}&
\includegraphics[width=9cm,angle=0]{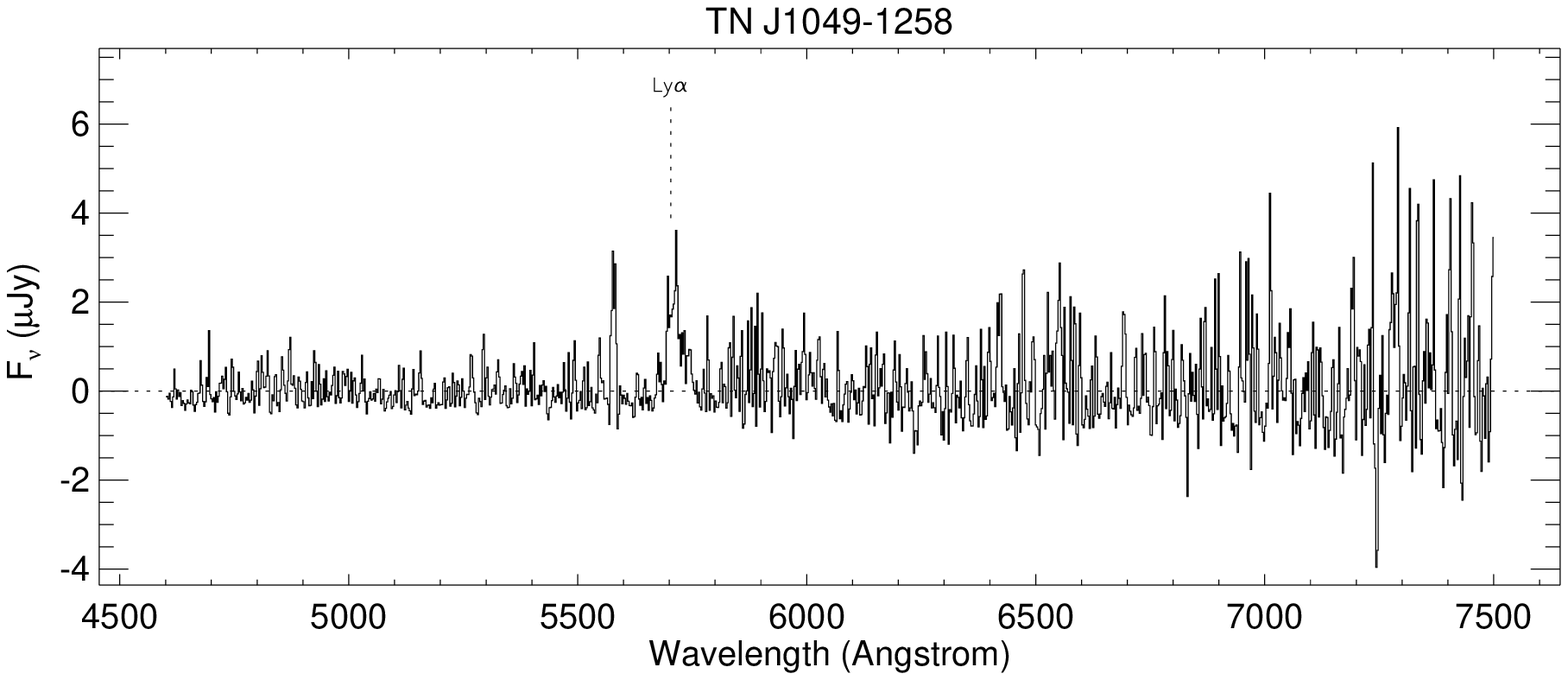}\\
\includegraphics[width=9cm,angle=0]{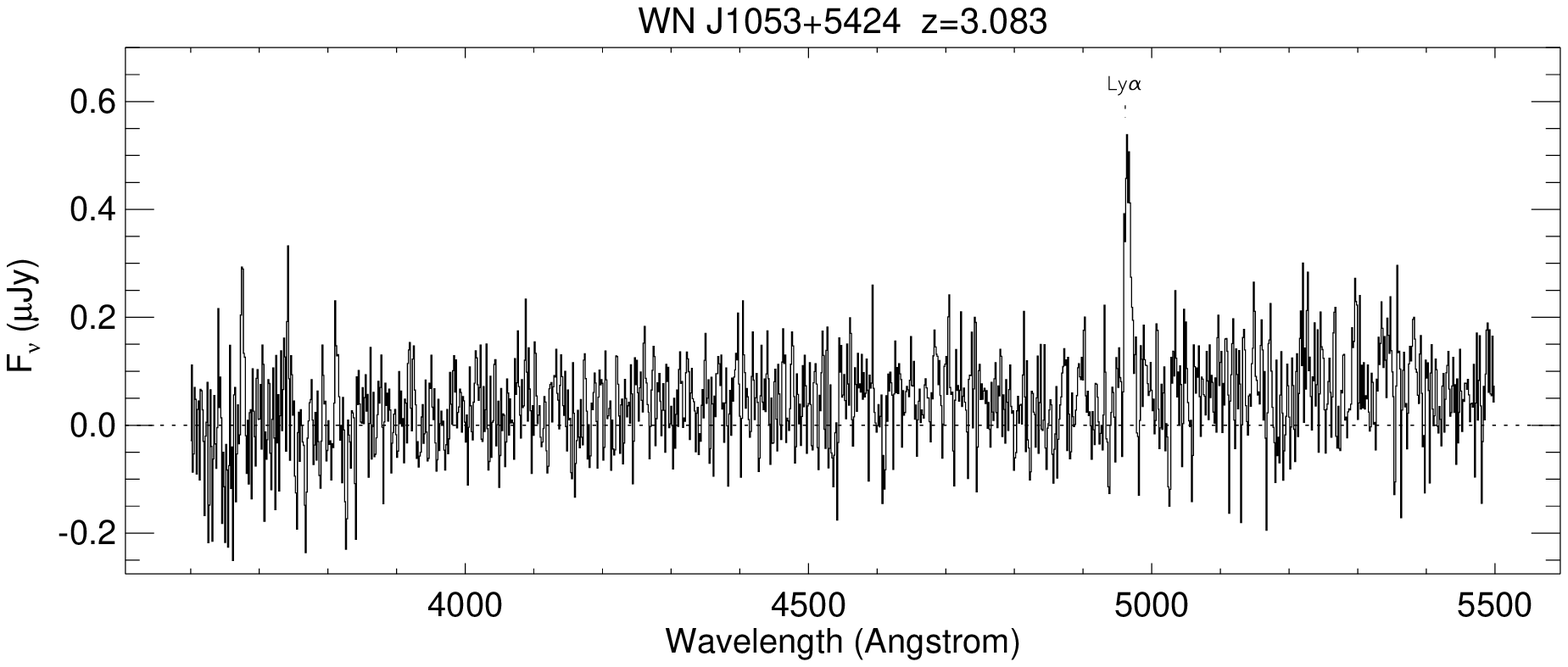}& 
\includegraphics[width=9cm,angle=0]{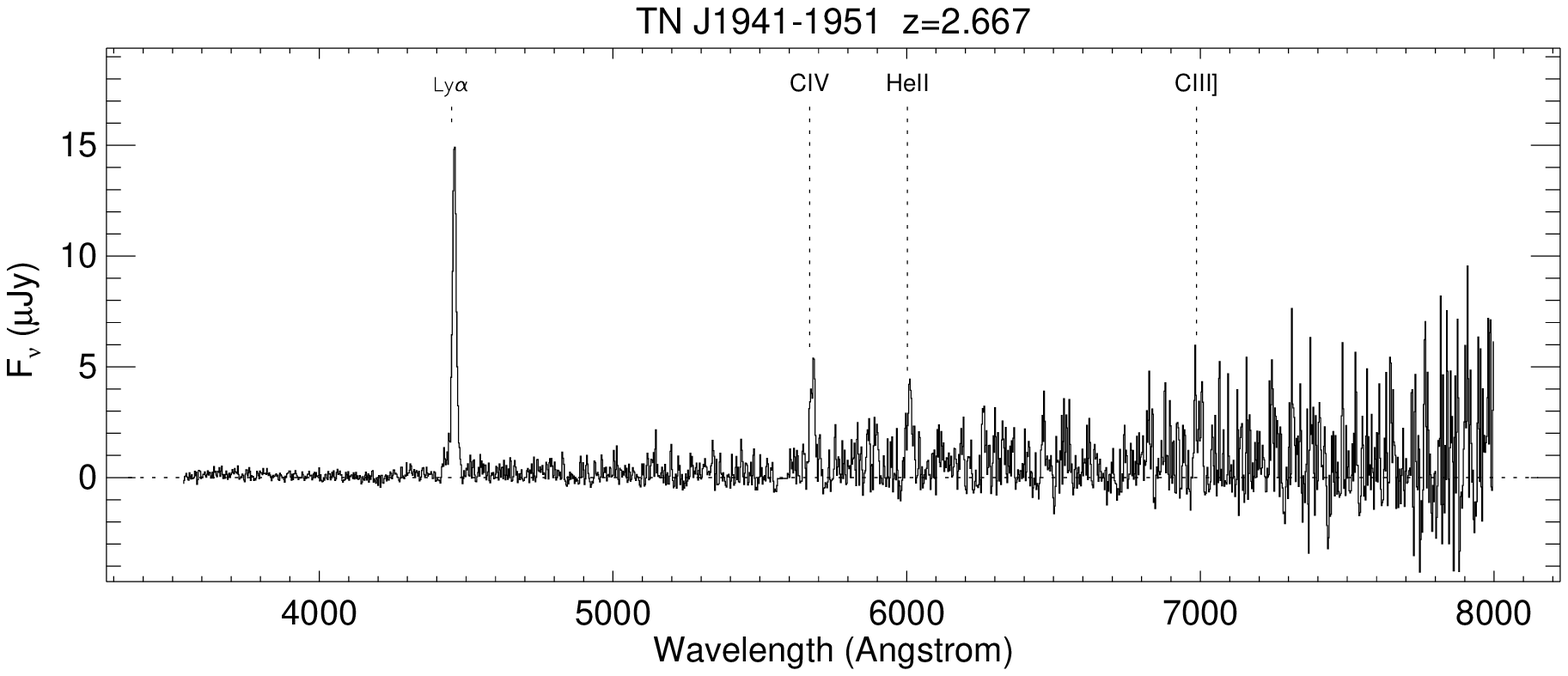}\\
\includegraphics[width=9cm,angle=0]{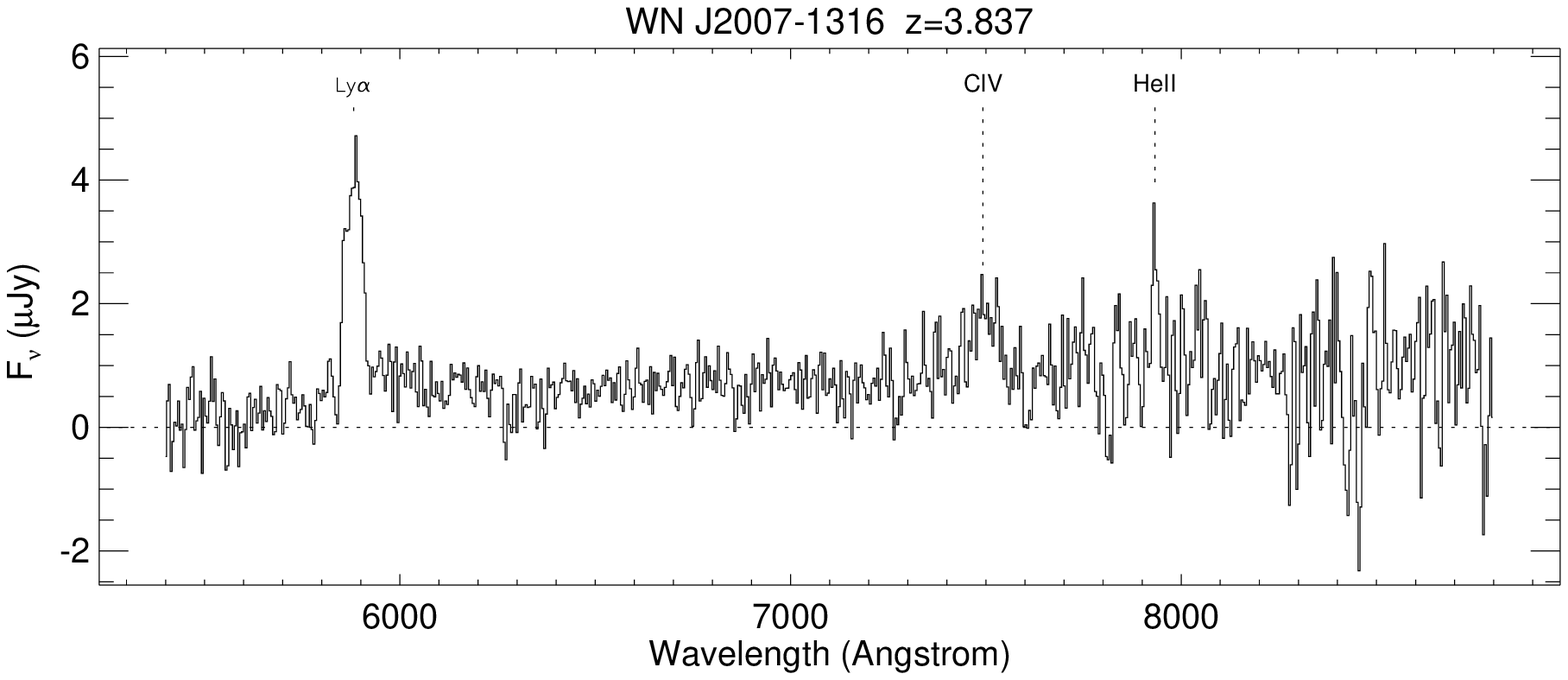}& 
\includegraphics[width=9cm,angle=0]{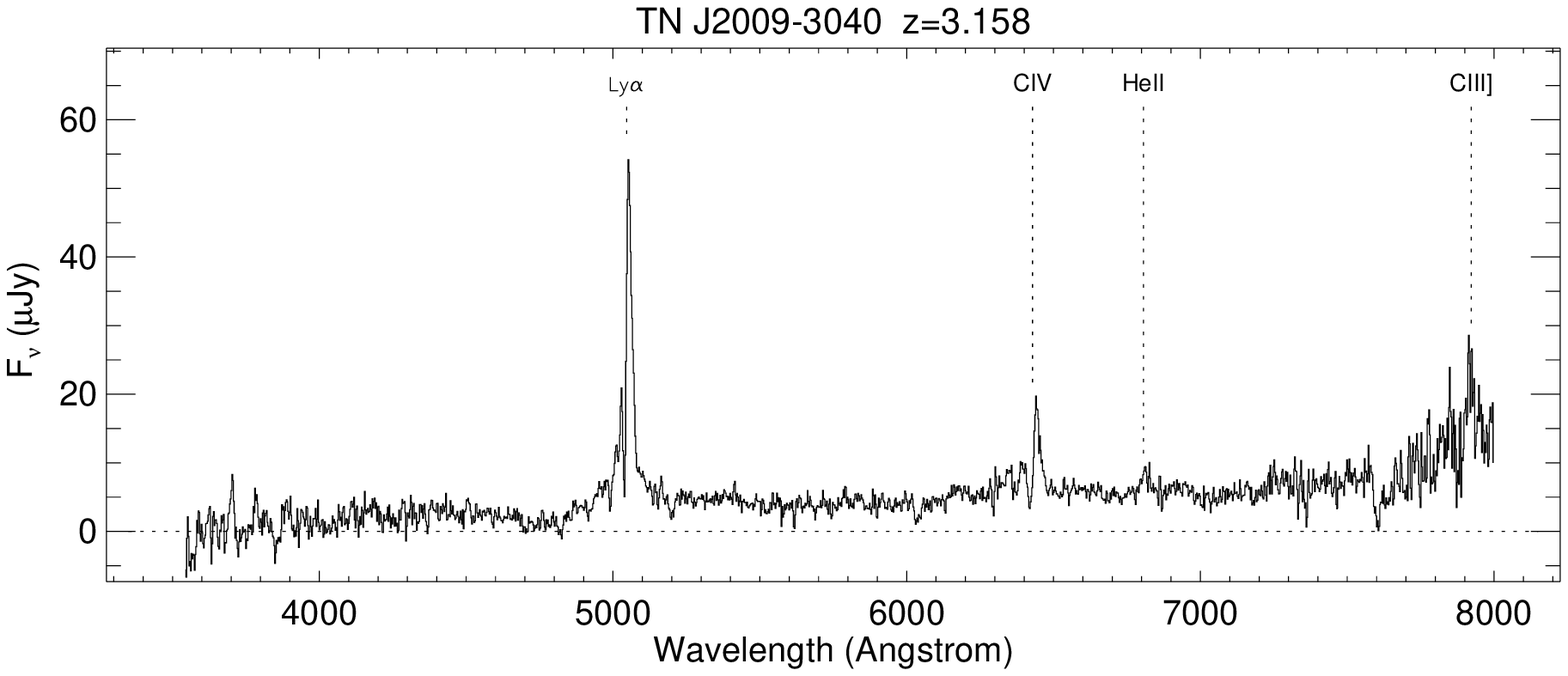}\\
\includegraphics[width=9cm,angle=0]{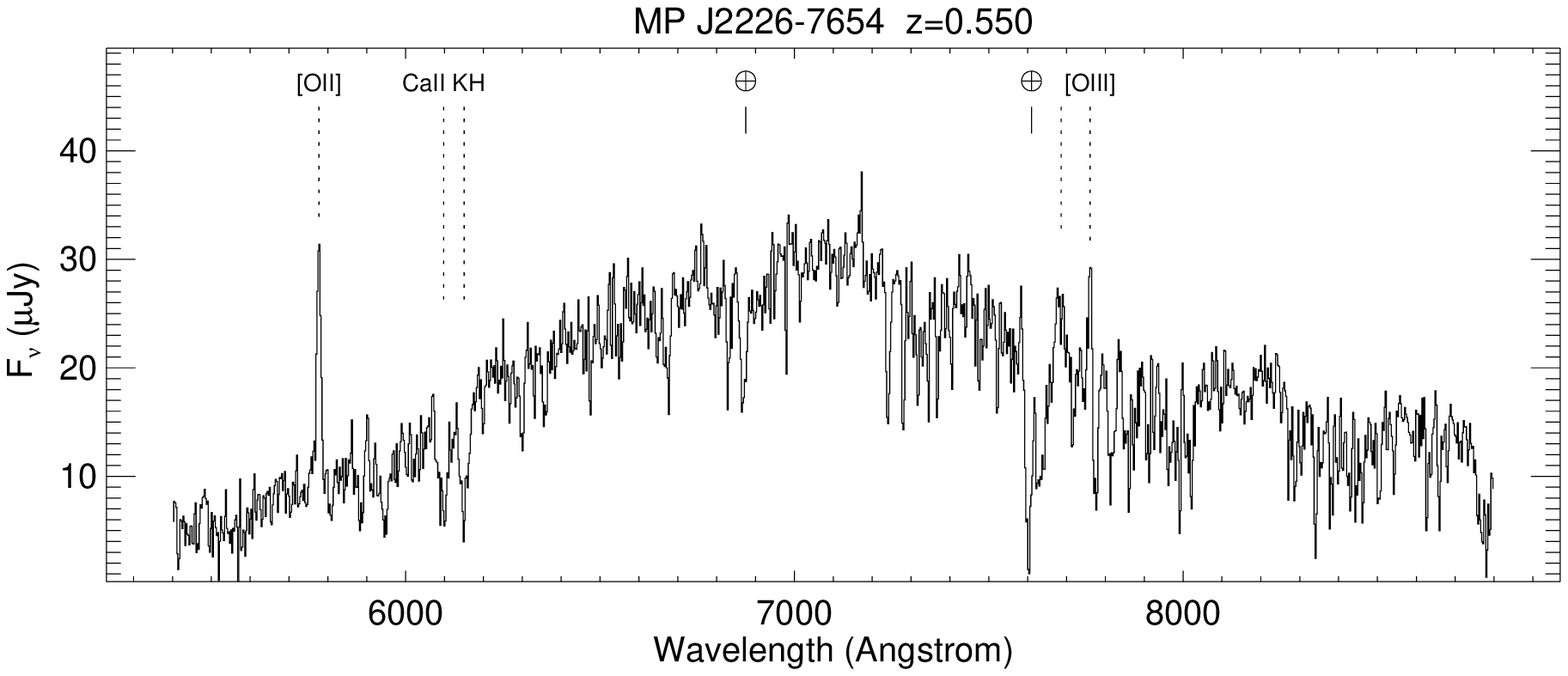}& 
\includegraphics[width=9cm,angle=0]{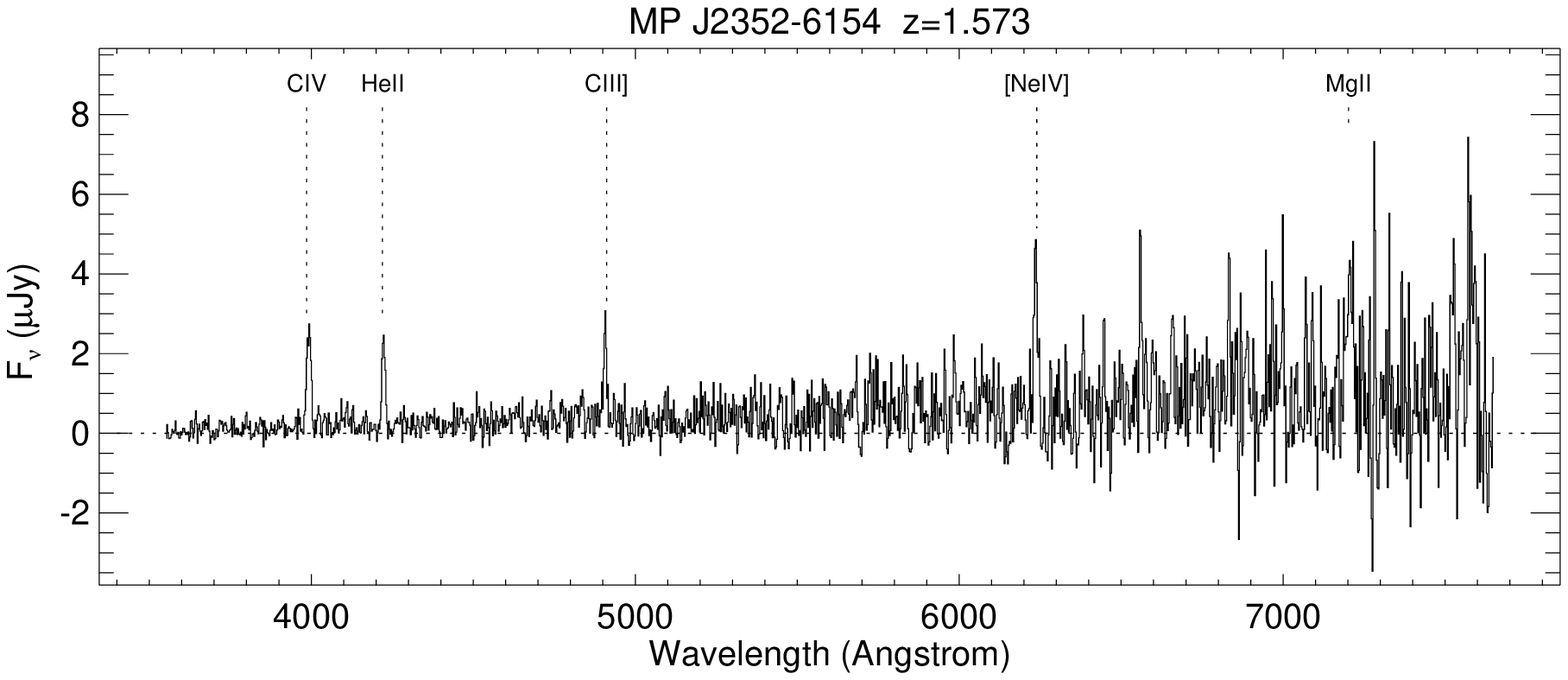}\\
\label{spec}
\end{tabular}
\caption{Optical spectroscopy of USS sources. Prominent features are indicated with vertical dotted lines showing the predicted observed wavelength of the lines at the redshift the source. Atmospheric absorption features are indicated with $\oplus$.
 }
\end{figure*}

\section{Acknowledgments}
We are grateful to the anonymous referee for his/her careful reading of the manuscript and a number of comments, which improved the paper.
We thank Adam Stanford and Cyril Tasse for their help in taking the Keck and WHT spectra, respectively.
This work was partially supported by the
Consejo Nacional de Investigaciones Cient\'{\i}ficas y T\'ecnicas (CONICET),
the Secretar\'ia de Ciencia y
T\'ecnica de la Universidad Nacional de C\'ordoba and the European Union Alfa II Programme, 
through LENAC, the Latin American--European Network for Astrophysics and Cosmology.
The work of WdV, WvB and SDC was performed under the auspices of the
U.S. Department of Energy, National Nuclear Security Administration by
the University of California, Lawrence Livermore National Laboratory
under contract No. W-7405-Eng-48.
Based on observations made with the European Southern Observatory
telescopes obtained from the ESO/ST-ECF Science Archive Facility.
The authors made use of the database CATS (Verkhodanov et al. 1997) of the Special Astrophysical Observatory 
and the 
NASA/IPAC extragalactic database (NED) which is operated by the Jet
Propulsion Laboratory, Caltech, under contract with the National
Aeronautics and Space Administration.
This publication makes use of data products
from the Two Micron All Sky Survey, which is a joint project of the
University of Massachusetts and the Infrared Processing and Analysis
Center/California Institute of Technology, funded by the National
Aeronautics and Space Administration and the National Science
Foundation.  

{}

\begin{table*}
{\bf Table 1.} Log of the $K$--band observations of the USS WISH sample.\\
\begin{center}
\begin{tabular}{lccc}
\hline
Name & Dates observed &Telescope/Instr.& Total exposure times (s)\\
\hline
WN~J0037$-$1904 &2001-01-09&\tiny{CTIO/CIRIM} & 1920  \\
WN~J0141$-$1406 &1999-07-20&\tiny{CTIO/CIRIM}& 1500  \\
WN~J0224$-$1701 &2002-07-07&\tiny{VLT/ISAAC}  &2340   \\
WN~J0246$-$1649 &2001-01-09&\tiny{CTIO/OSIRIS}&1920 \\
WN~J0510$-$1838 &2001-01-09&\tiny{CTIO/OSIRIS}&960   \\
WN~J0604$-$2015 &2002-09-02&\tiny{VLT/ISAAC  }&2340\\
WN~J0910$-$2228 &2000-03-22&\tiny{CTIO/OSIRIS}&3840\\
WN~J0912$-$1655 &2000-03-21&\tiny{CTIO/OSIRIS}&1920\\
WN~J1047$-$1836 &2000-03-22&\tiny{CTIO/OSIRIS}&3840\\
WN~J1052$-$1812 &2000-03-23&\tiny{CTIO/OSIRIS}&1920\\
WN~J1101$-$2134 &2002-04-23&\tiny{VLT/ISAAC}&1440\\
WN~J1109$-$1917 &2000-03-23&\tiny{CTIO/OSIRIS}&1920\\
WN~J1132$-$2102 &2001-01-09&\tiny{CTIO/OSIRIS}&3720\\
WN~J1138$-$1324 &2002-04-23&\tiny{VLT/ISAAC}  &2340\\
WN~J1150$-$1317 &2000-03-22&\tiny{CTIO/OSIRIS}&3840\\
WN~J1222$-$2129 &2000-03-21&\tiny{CTIO/OSIRIS}&3360\\
WN~J1255$-$1913 &2000-03-23&\tiny{CTIO/OSIRIS}&1920\\
WN~J1331$-$1947 &2000-03-23&\tiny{CTIO/OSIRIS}&1920\\
WN~J1450$-$1525 &2000-03-21&\tiny{CTIO/OSIRIS}&2160\\
WN~J1516$-$2110 &2002-04-23&\tiny{VLT/ISAAC}  &2340\\
WN~J1518$-$1225 &2000-03-23&\tiny{CTIO/OSIRIS}&3840\\
WN~J1557$-$1349 &2000-03-22&\tiny{CTIO/OSIRIS}&1920\\
WN~J1603$-$1500 &2002-04-23&\tiny{VLT/ISAAC}  &2340\\
WN~J1637$-$1931 &2000-03-21&\tiny{CTIO/OSIRIS}&1280\\
WN~J2002$-$1842 &2002-04-23&\tiny{VLT/ISAAC}  &2340\\
WN~J2007$-$1316 &1999-07-20&\tiny{CTIO/CIRIM} &3840   \\
WN~J2214$-$2353 &1999-07-20&\tiny{CTIO/CIRIM} &2880   \\
WN~J2343$-$1947 &1999-07-20&\tiny{CTIO/CIRIM} &1920   \\
\hline
\end{tabular}
\end{center}
\end{table*}

\begin{onecolumn}
\begin{sidewaystable}\small
\begin{minipage}{225mm}
{\bf Table 2.} Results of the $K-$band and radio observations of the WISH--NVSS USS sample\\
\begin{center}
\begin{small}
\begin{tabular}{lrrrrrrrrrrrr}
\hline
Name & $\alpha_{352}^{1400}$&LAS &K(2\arcsec) & K(4\arcsec)&K(8\arcsec)& $K_{\tt BEST}$ &  $RA^{radio}_{J2000}$ & $DEC^{radio}_{J2000}$ & $RA^{K-band}_{J2000}$& $DEC^{K-band}_{J2000}$ &$r_{1/2}$&{\tt S/G}  \\
& &($\arcsec$)&(mag)&(mag)&(mag)&(mag)&$^h\;\; ^m\;\;\;\; ^s\;\;\,$ & \degr$\;\;\;$ \arcmin$\;\;\;$ \arcsec$\;$ &$^h\;\; ^m\;\;\;\; ^s\;\;\,$ & \degr$\;\;\;$ \arcmin$\;\;\;$ \arcsec$\;$&(\arcsec) &\\
\hline
WN~J0037$-$1904 &   $-$1.36$\pm$0.04 &  14.9 &19.05$\pm$0.09  &18.30$\pm$0.09  &17.56$\pm$0.09   &17.38$\pm$0.09   &00 37 23.8  &$-$19 04 35.6 &00 37 23.79 &$-$ 19 04 38.37 & 0.36  & 0.02   \\
WN~J0141$-$1406 &   $-$1.68$\pm$0.05 &  15.7 &19.54$\pm$0.19  &18.71$\pm$0.18  &18.80$\pm$0.33   &18.50$\pm$0.22   &01 41 38.0  &$-$14 06 09.7 &01 41 37.94 &$-$ 14 06 09.17 & 0.25  & 0.11   \\
WN~J0224$-$1701 &   $-$1.36$\pm$0.08 &  8.2  &\nodata         &$>$22.0         &\nodata          &\nodata          &02 24 15.7  &$-$17 01 10.3 &\nodata     & \nodata        &\nodata&\nodata \\
WN~J0246$-$1649 &   $-$1.38$\pm$0.04 &$<$4.4 &\nodata         &$>$23.0         &\nodata          &\nodata          &02 46 52.9  &$-$16 49 28.7 &\nodata     &\nodata         &\nodata&\nodata \\
WN~J0510$-$1838$^*$&$-$1.65$\pm$0.04 &  36.1 &15.63$\pm$0.01  &14.90$\pm$0.01  &14.35$\pm$0.01   &14.05$\pm$0.10   &05 10 32.2  &$-$18 38 41.4 &05 10 31.96 &$-$18 38 44.00  &2.62   & 0.03   \\
WN~J0604$-$2015 &   $-$1.51$\pm$0.04 &  11.2 &20.23$\pm$0.06  &19.63$\pm$0.06  &19.14$\pm$0.08   &19.54$\pm$0.08   &06 04 57.4  &$-$20 15 56.5 &06 04 57.71 &$-$20 15 57.01  &0.95   & 0.00   \\
WN~J0910$-$2228$^*$&$-$1.57$\pm$0.04 &$<$2.0 &19.35$\pm$0.16  &18.97$\pm$0.21  &19.04$\pm$0.37   &18.93$\pm$0.17   &09 10 34.1  &$-$22 28 47.4 &09 10 34.15 &$-$22 28 47.13  &0.72   & 0.00   \\
WN~J0912$-$1655 &   $-$1.58$\pm$0.06 &$<$5.0 &18.18$\pm$0.11  &17.69$\pm$0.12  &17.23$\pm$0.14   &17.51$\pm$0.13   &09 12 57.1  &$-$16 55 55.5 &09 12 57.17 &$-$16 55 55.58  &0.81   & 0.08   \\
WN~J1047$-$1836 &   $-$1.46$\pm$0.05 &$<$4.5 &18.94$\pm$0.13  &18.54$\pm$0.15  &18.28$\pm$0.21   &18.58$\pm$0.15   &10 47 15.5  &$-$18 36 30.6 &10 47 15.39 &$-$18 36 30.32  &0.64   & 0.02   \\
WN~J1052$-$1812 &   $-$1.52$\pm$0.05 &  11.7 &17.45$\pm$0.08  &16.91$\pm$0.09  &16.65$\pm$0.10   &16.98$\pm$0.08   &10 52 00.8  &$-$18 12 31.4 &10 52 00.88 &$-$18 12 30.08  &0.98   & 0.01   \\
WN~J1101$-$2134 &   $-$1.67$\pm$0.07 &$<$4.4 &\nodata         &$>$21.0         &\nodata          &\nodata          &11 01 54.2  &$-$21 34 28.4 &\nodata     &\nodata         &\nodata&\nodata \\
WN~J1109$-$1917 &   $-$1.38$\pm$0.04 &$<$4.4 &20.27$\pm$0.41  &20.76$\pm$1.17  &\nodata          &20.26$\pm$0.37   &11 09 49.9  &$-$19 17 54.0 &11 09 49.89 &$-$19 17 53.66  &0.34   & 0.39   \\
WN~J1132$-$2102 &   $-$1.42$\pm$0.04 &$<$4.4 &\nodata         &$>$23.0         &\nodata          &\nodata          &11 32 52.7  &$-$21 02 44.7 &\nodata     & \nodata        &\nodata&\nodata \\
WN~J1138$-$1324 &   $-$1.53$\pm$0.08 &$<$4.6 &\nodata         &$>$22.0         &\nodata          &\nodata          &11 38 05.0  &$-$13 24 22.7 &\nodata     &\nodata         &\nodata&\nodata \\
WN~J1150$-$1317 &   $-$1.37$\pm$0.04 &$<$4.6 &19.32$\pm$0.15  &18.94$\pm$0.19  &18.37$\pm$0.22   &18.93$\pm$0.19   &11 50 09.6  &$-$13 17 54.1 &11 50 09.52 &$-$13 17 52.82  &0.68   &0.00    \\
WN~J1222$-$2129 &   $-$1.42$\pm$0.06 &$<$4.5 &\nodata         &$>$23.0         &\nodata          &\nodata          &12 22 48.2  &$-$21 29 11.1 &\nodata     &\nodata         &\nodata&\nodata \\
WN~J1255$-$1913 &   $-$1.67$\pm$0.06 &$<$4.5 &\nodata         &$>$23.0         &\nodata          &\nodata          &12 55 52.4  &$-$19 13 01.5 &\nodata     &\nodata         &\nodata&\nodata \\
WN~J1331$-$1947 &   $-$1.40$\pm$0.04 &$<$6.4 &18.15$\pm$0.11  &17.86$\pm$0.13  &17.86$\pm$0.21   &17.88$\pm$0.13   &13 31 47.2  &$-$19 47 26.7 &13 31 47.18 &$-$19 47 25.87  &0.57   & 0.52   \\
WN~J1450$-$1525 &   $-$1.42$\pm$0.11 &$<$5.2 &23.71$\pm$0.10  &23.08$\pm$0.10  &22.85$\pm$0.13   &22.77$\pm$0.11   &14 50 42.7  &$-$15 25 39.2 &14 50 42.58 &$-$15 25 40.51  &1.14   & 0.04   \\
WN~J1516$-$2110 &   $-$1.38$\pm$0.04 &$<$5.4 &\nodata         &$>$22.0         &\nodata          &\nodata          &15 16 42.4  &$-$21 10 27.7 &\nodata     &\nodata         &\nodata&\nodata \\
WN~J1518$-$1225 &   $-$1.67$\pm$0.06 &$<$5.1 &20.25$\pm$0.20  &22.08$\pm$0.41  &\nodata          &20.21$\pm$0.34   &15 18 43.8  &$-$12 25 30.6 &15 18 43.48 &$-$12 25 34.36  &0.44   & 0.67   \\
WN~J1557$-$1349 &   $-$1.39$\pm$0.06 &$<$5.3 &18.64$\pm$0.13  &18.21$\pm$0.16  &18.59$\pm$0.35   &18.32$\pm$0.17   &15 57 41.5  &$-$13 49 59.1 &15 57 41.61 &$-$13 49 57.63  &0.70   & 0.03   \\
WN~J1603$-$1500 &   $-$1.44$\pm$0.05 &$<$5.3 &20.08$\pm$0.07  &20.05$\pm$0.14  &20.11$\pm$0.31   &19.99$\pm$0.10   &16 03 04.6  &$-$15 00 53.0 &16 03 04.65 &$-$15 00 53.68  &0.52   &0.69    \\
WN~J1637$-$1931 &   $-$1.60$\pm$0.04 &$<$5.5 &\nodata         &$>$22.0         &\nodata          &\nodata          &16 37 44.8  &$-$19 31 24.2 &\nodata     &\nodata         &\nodata&\nodata \\
WN~J2002$-$1842 &   $-$1.42$\pm$0.06 &$<$7.6 &20.42$\pm$0.10  &20.10$\pm$0.16  &21.07$\pm$0.77   &20.21$\pm$0.13   &20 02 56.0  &$-$18 42 46.3 &20 02 55.99 &$-$18 42 46.25  &0.58 &0.22      \\
WN~J2007$-$1316 &   $-$1.52$\pm$0.04 &$<$7.4 &18.75$\pm$0.05  &18.36$\pm$0.07  &18.08$\pm$0.09   &18.25$\pm$0.07   &20 07 53.2  &$-$13 16 43.7 &20 07 53.22 &$-$13 16 43.70  &0.77 &0.57      \\
WN~J2214$-$2353 &   $-$1.34$\pm$0.05 &  12.7 &18.62$\pm$0.05  &17.90$\pm$0.05  &17.37$\pm$0.06   &17.64$\pm$0.05   &22 14 14.0  &$-$23 53 24.8 &22 14 14.17 &$-$23 53 27.46  &1.63 &0.03      \\
WN~J2343$-$1947 &   $-$1.80$\pm$0.04 &$<$6.4 &21.07$\pm$0.19  &20.48$\pm$0.22  &20.42$\pm$0.39   &20.09$\pm$0.22   &23 43 16.3  &$-$19 47 15.4 &23 43 16.23 &$-$19 47 15.83  &1.13 &0.00      \\
\hline
\end{tabular}
\end{small}
\begin{flushleft}
$^*$ ATCA image \citep{DB00}
\end{flushleft}
\end{center}
\end{minipage}
\end{sidewaystable}
\end{onecolumn}
\twocolumn

\begin{table*}
{\bf Table 3.} Log of the Keck I optical spectroscopy.\\
\begin{center}
\begin{small}
\begin{tabular}{lcrccrcccc}
\hline
Target & Dichroic  & Blue grism & Blue pix. scale & Resolution & Red grating & Red pix. scale  & Resolution & Exp. time  & PA \\
\hline
WN\,J0528+6549 & D680& 400/3400   & 1.04 \AA\ / pix & 9.5\AA & 400/8500 & 1.86\AA\ / pix & 8.1\AA  & $1 \times 1800$\,s & 177.00 \\
WN\,J0716+5107 & D680 &400/3400   & 1.04 \AA\ / pix & 9.5\AA & 400/8500 & 1.86\AA\ / pix & 8.1\AA  & $2 \times 1800$\,s & 201.00 \\
WN\,J1053+5424 & D560& 300/5000   & 1.43 \AA\ / pix &  12\AA & 400/8500 & 1.86\AA\ / pix & 8.1\AA  & $4 \times 1800$\,s & 254.06 \\

\hline
\end{tabular}
\end{small}
\end{center}
\end{table*}

\begin{table*}
{\bf Table 4.} Emission line measurements
\begin{center}
\begin{tabular}{lrrrrrr}
\hline
\hline
\multicolumn{1}{c}{Source} & \multicolumn{1}{c}{$z$} & \multicolumn{1}{c}{Line} & \multicolumn{1}{c}{$\lambda_{\rm 
obs}$} & 
\multicolumn{1}{c}{Flux} & \multicolumn{1}{c}{$\Delta v_{\rm FWHM}$} & \multicolumn{1}{c}{$W_{\lambda}^{\rm 
rest}$}\\[1mm]
 &  &  & \multicolumn{1}{c}{\AA} & $10^{-16}$ erg\,s$^{-1}$\,cm$^{-2}$ & km s$^{-1}$ & \multicolumn{1}{c}{\AA} \\
\hline
\hline
MP~J0340$-$6507       & 2.289$\pm$0.005 & \Lya  & 4000$\pm$4  & 0.40$\pm$0.05 & 1860$\pm$640  & $>$65\\
                      &                 & \CIV  & 5100$\pm$27 & 0.15$\pm$0.05 & 2300$\pm$1800 & 20$\pm$10 \\
                      &                 & \HeII & 6275$\pm$10 & 0.26$\pm$0.05 & 1400$\pm$500  & $>$40 \\
                      &                 & \CIII & 6270$\pm$10 & 0.25$\pm$0.05 & 1600$\pm$1100 & $>$130\\ 
\hline
WN~J0528$+$6549       & 1.217$\pm$0.005 & \OII  &  8263$\pm$17 & 0.09$\pm$0.02 & 1400$\pm$900 & $>$31 \\
\hline
MP~J0601$-$3926       & 1.633$\pm$0.001 & \CIV  & 4083$\pm$27 & 0.30$\pm$0.15 & 2300$\pm$1600 & $>$4 \\
                      &                 & \CIII & 5020$\pm$4  & 0.6$\pm$0.1   & 1200$\pm$500  & 28$\pm$6 \\
                      &                 & \CII  & 6130$\pm$4 &  0.20$\pm$0.04 & 550$\pm$360 & 20$\pm$7 \\
\hline
WN~J0716$+$5107      &1.1401$\pm$0.0002 & \OII & 7976$\pm$1 & 0.06$\pm$0.01 & $<$500 & 8$\pm$1 \\
\hline
WN~J0912$-$1655      & 0.9102$\pm$0.0002 & \OII & 7119$\pm$1 & 0.10$\pm$0.02 & $<$100 & 11$\pm$2 \\
\hline
TN~J1049$-$1258       & 3.697$\pm$0.004 & \Lya  & 5712$\pm$5 & 0.6$\pm$0.1    & 1600$\pm$500  &$>$63 \\
\hline
WN~J1053$+$5424       & 3.083$\pm$0.001 & \Lya  & 4965$\pm$1 & 0.06$\pm$0.01    & $<$500  &25$\pm$6 \\
\hline
TN~J1941$-$1951       & 2.667$\pm$0.001 & \Lya  & 4460$\pm$1 & 3.7$\pm$0.4 & 800$\pm$200 & 330$\pm$70 \\
                      &                 & \CIV  & 5680$\pm$2 & 0.9$\pm$0.1 & 850$\pm$200 &  64$\pm$20 \\
                      &                 & \HeII & 6010$\pm$3 & 0.4$\pm$0.1 & 850$\pm$300 & 23$\pm$6 \\
                      &                 & \CIII & 7000$\pm$16& 0.2$\pm$0.1 & 1000$\pm$700 & 10$\pm$4 \\
\hline
WN~J2007$-$1316       & 3.837$\pm$0.001 & \Lya  & 5883$\pm$3 & 1.8$\pm$0.2 &2400$\pm$300 & 77$\pm$13 \\
                      &                 & \CIV  & 7500$\pm$20 & 0.5$\pm$0.1 & 3100$\pm$1700 &  26$\pm$5 \\
                      &                 & \HeII & 7933$\pm$2 & 0.2$\pm$0.1 & 450$\pm$200 & 9$\pm$2 \\
\hline

TN~J2009$-$3040       & 3.158$\pm$0.001 & \Lya & 5056$\pm$1 & 15.8$\pm$1.6 & 2300$\pm$500$\dagger$ & 65$\pm$7 \\
                      &                 & \CIV & 6444$\pm$1 & 2.0$\pm$2.0  & 2300$\pm$500$\dagger$ & 10$\pm$1 \\
                      &                 & \HeII & 6815$\pm$5 & 0.65$\pm$0.1  & 1380$\pm$500 & 4$\pm$1 \\
                      &                 & \CIII & 7920$\pm$5 & 3.0$\pm$0.4  & 2900$\pm$600$\dagger$ & 14$\pm$2 \\

\hline
MP~J2226$-$7654       &0.5500$\pm$0.0002& \OII & 5776$\pm$1 & 2.8$\pm$0.3 & 400$\pm$180 & 25$\pm$3 \\
                      &                 &CaII K& 6098$\pm$1& $-$2.3$\pm$0.3& 1200$\pm$300 & -12$\pm$2  \\
                      &                 &CaII H& 6161$\pm$1& $-$2.0$\pm$0.3& 1000$\pm$200 & -10$\pm$2  \\

                      &                 &\OIII & 7680$\pm$10& 1.2$\pm$0.2& 900$\pm$500 & 4$\pm$1       \\
                      &                 &\OIII & 7760$\pm$5& 0.8$\pm$0.2& 800$\pm$600 & 8$\pm$2       \\
\hline
MP~J2352$-$6154       &1.573$\pm$0.003  & \CIV & 3990$\pm$1 & 0.7$\pm$0.1 & 800$\pm$260 & 80$\pm$16 \\
                      &                 & \HeII& 4223$\pm$1& 0.40$\pm$0.04& $<$500 & 40$\pm$7    \\
                      &                 & \CIII& 4900$\pm$1& 0.30$\pm$0.04& $<$500 & 31$\pm$5    \\
                      &                 & \NeIV& 6235$\pm$1& 0.40$\pm$0.05& 340$\pm$200 & 30$\pm$10    \\
                      &                 & \MgII& 7200$\pm$4& 0.3$\pm$0.1& 600$\pm$400 & 30$\pm$10    \\
\hline
\hline
\end{tabular}
\end{center}
\begin{flushleft}
$\dagger$ Not considering the broad underlying component.
\end{flushleft}
\end{table*}


\begin{thebibliography}{}

\bibitem[Bertin \& Arnouts(1996)]{bertin} {Bertin} E., {Arnouts} S.,1996, A \& A Supp. 117, 393.

\bibitem[Best et al.(1998)]{best98} Best, P.~N., Longair, 
M.~S., \& Roettgering, H.~J.~A.\ 1998, \mnras, 295, 549 

\bibitem[Bornancini et al.(2004)]{bornan04} Bornancini, C.~G., 
Mart{\'{\i}}nez, H.~J., Lambas, D.~G., de Vries, W., van Breugel, W., De 
Breuck, C., \& Minniti, D.\ 2004, \aj, 127, 679 

\bibitem[Bornancini et al.(2006)]{bornan06} Bornancini, C.~G., 
Lambas, D.~G., \& De Breuck, C.\ 2006, \mnras, 366, 1067 

\bibitem[Bornancini et al.(2006)]{bornan} Bornancini, C.~G., 
Padilla, N.~D., Lambas, D.~G., \& De Breuck, C.\ 2006, \mnras, 368, 619 


\bibitem[Blundell et al.(1998)]{blundell} Blundell, K.~M., 
Rawlings, S., Eales, S.~A., Taylor, G.~B., \& Bradley, A.~D.\ 1998, \mnras, 
295, 265 

\bibitem[Carilli et al.(1999)]{carilli} Carilli, C.~L., 
R{\"o}ttgering, H.~J.~A., Miley, G.~K., Pentericci, L.~H., \& Harris, 
D.~E.\ 1999, The Most Distant Radio Galaxies, 123 

\bibitem[Carter \etal(1994)]{car94} Carter, D. et al. 1994, ISIS Users' Manual

\bibitem[Cohen et al.(2006)]{cohen} Cohen, A.~S., Lane, 
W.~M., Kassim, N.~E., Lazio, T.~J.~W., Cotton, W.~D., Perley, R.~A., 
Condon, J.~J., \& Erickson, W.~C.\ 2006, Astronomische Nachrichten, 327, 
262 

\bibitem[Cruz et al.(2006)]{cruz} Cruz, M.~J., et al.\ 
2006, \mnras, 373, 1531

\bibitem[Cruz et al.(2007)]{cruz07} Cruz, M.~J., et al.\ 
2007, \mnras, in press, astro-ph/0612268

\bibitem[De Breuck et al.(2000)]{DB00} De Breuck, C., van 
Breugel, W., R{\"o}ttgering, H.~J.~A., \& Miley, G.\ 2000, \aaps, 143, 303 

\bibitem[De Breuck et al.(2001)]{DB01} De Breuck, C., et al.\ 2001, \aj, 121, 1241 

\bibitem[De Breuck et al.(2002a)]{wish} De Breuck, C., Tang, 
Y., de Bruyn, A.~G., R{\"o}ttgering, H., \& van Breugel, W.\ 2002b, \aap, 
394, 59 

\bibitem[De Breuck et al.(2002b)]{DB02} De Breuck, C., van Breugel, W., Stanford, S.~A., R{\"o}ttgering, H., Miley, G., \& Stern, D.\ 2002a, \aj, 123, 637 

\bibitem[De Breuck et al.(2004)]{sumss} De Breuck, C., 
Hunstead, R.~W., Sadler, E.~M., Rocca-Volmerange, B., \& Klamer, I.\ 2004, 
\mnras, 347, 837 

\bibitem[De Breuck et al.(2006)]{DB06} De Breuck, C., 
Klamer, I., Johnston, H., Hunstead, R.~W., Bryant, J., Rocca-Volmerange, 
B., \& Sadler, E.~M.\ 2006, \mnras, 366, 58 

\bibitem[Cutri et al.(2003)]{cutri}Cutri, R. M., Skrutskie, M. F., van Dyk, S., et al. 2003, The 2MASS All-Sky Catalogue of Point Sources

\bibitem[Eales \& Rawlings(1993)]{eales} Eales, S.~A., \& 
Rawlings, S.\ 1993, \apj, 411, 67 

\bibitem[Eales et al.(1997)]{eales97} Eales, S., Rawlings, S., 
Law-Green, D., Cotter, G., \& Lacy, M.\ 1997, \mnras, 291, 593 

\bibitem[Gopal-Krishna(1988)]{gopal} Gopal-Krishna 1988, 
\aap, 192, 37

\bibitem[Jarvis et al.(2001)]{jarvis01} Jarvis, M.~J., Rawlings, 
S., Eales, S., Blundell, K.~M., Bunker, A.~J., Croft, S., McLure, R.~J., \& 
Willott, C.~J.\ 2001, \mnras, 326, 1585 

\bibitem[Jarvis et al.(2004)]{jarvis04} Jarvis, M.~J., Cruz, 
M.~J., Cohen, A.~S., R{\"o}ttgering, H.~J.~A., \& Kassim, N.~E.\ 2004, 
\mnras, 355, 20 

\bibitem[Klamer et al.(2006)]{klamer} Klamer, I.~J., Ekers, 
R.~D., Bryant, J.~J., Hunstead, R.~W., Sadler, E.~M., \& de Breuck, C.\ 
2006, \mnras, 371, 852 

\bibitem[Krolik \& Chen(1991)]{krolik} Krolik, J.~H., \& Chen, 
W.\ 1991, \aj, 102, 1659 

\bibitem[Lacy et al.(2000)]{lacy} Lacy, M., Bunker, A.~J., 
\& Ridgway, S.~E.\ 2000, \aj, 120, 68 

\bibitem[Miley et al.(2004)]{mileynat} Miley, G.~K., et al.\ 
2004, \nat, 427, 47 

\bibitem[Miley et al.(2006)]{miley} Miley, G.~K., et al.\ 
2006, \apjl, 650, L29 

\bibitem[Mink(2006)]{wcs} Mink, D.\ 2006, ASP 
Conf.~Ser.~351: Astronomical Data Analysis Software and Systems XV, 351, 
204 

\bibitem[Moorwood et al.(1998)]{moor} Moorwood, A., et al.\ 
1998, The Messenger, 94, 7 

\bibitem[Monet et al.(1998)]{monet} Monet, D.~B.~A., et al.\ 
1998, VizieR Online Data Catalog, 1252, 0 

\bibitem[Nesvadba et al.(2006)]{nesvadba} Nesvadba, N., Lehnert, M., Eisenhauer, F., Gilbert, A., Tecza, M., \& Abuter, R.\ 2006, \apj, 650, 693

\bibitem[Oke et al.(1995)]{lris} Oke, J.~B., et al.\ 1995, \pasp, 107, 375

\bibitem[Overzier et al.(2006)]{overzier} Overzier, R.~A., et 
al.\ 2006, \apj, submitted 

\bibitem[Pandey(2006)]{pandey} Pandey, V. 2006, PhD thesis, Raman Research Institute, Bangalore, India

\bibitem[Pedani(2003)]{pedani} Pedani, M.\ 2003, New 
Astronomy, 8, 805 

\bibitem[Persson et al.(1998)]{persson} Persson, S.~E., Murphy, 
D.~C., Krzeminski, W., Roth, M., \& Rieke, M.~J.\ 1998, \aj, 116, 2475 

\bibitem[Reuland et al.(2003a)]{reulanda} Reuland, M., et al.\ 
2003, \apj, 592, 755 

\bibitem[Reuland et al.(2003b)]{reulandb} Reuland, M., van 
Breugel, W., R{\"o}ttgering, H., de Vries, W., De Breuck, C., \& Stern, D.\ 
2003, \apjl, 582, L71 

\bibitem[Reuland et al.(2004)]{reuland04} Reuland, M., R{\"o}ttgering, H., van Breugel, W., \& De Breuck, C.\ 2004, \mnras, 353, 377 

\bibitem[Rocca-Volmerange et al.(2004)]{rocca} 
Rocca-Volmerange, B., Le Borgne, D., De Breuck, C., Fioc, M., \& Moy, E.\ 
2004, \aap, 415, 931 

\bibitem[R\"ottgering et al.(1997)]{roet97} R\"ottgering, 
H.~J.~A., van Ojik, R., Miley, G.~K., Chambers, K.~C., van Breugel, 
W.~J.~M., \& de Koff, S.\ 1997, \aap, 326, 505 

\bibitem[Rottgering et al.(2006)]{roet06} R\"ottgering, H., et al.\ 2006, in proceedings "Cosmology, galaxy formation and astroparticle physics on the pathway to the SKA", Oxford, astro-ph/0610596 

\bibitem[Schlegel et al.(1998)]{sch} Schlegel, D.~J., 
Finkbeiner, D.~P., \& Davis, M.\ 1998, \apj, 500, 525 

\bibitem[Seymour et al.(2006)]{seymour}Seymour et al.\ 2007 ApJS, in press, astro-ph/0703324

\bibitem[Stanford et al.(1995)]{stan} Stanford, S.~A., 
Eisenhardt, P.~R.~M., \& Dickinson, M.\ 1995, \apj, 450, 512 

\bibitem[van Breugel et al.(2006)]{breugel06} van Breugel, W., de 
Vries, W., Croft, S., De Breuck, C., Dopita, M., Miley, G., Reuland, M., R{\"o}ttgering 
, H.\ 2006, Astronomische Nachrichten, 327, 175 

\bibitem[Venemans et al.(2006)]{vene} Venemans, B.~P.et al.\ 2006, 
\aap~accepted  

\bibitem[Verkhodanov et al.(1997)]{verkho} Verkhodanov, O.~V., 
Trushkin, S.~A., \& Chernenkov, V.~N.\ 1997, Baltic Astronomy, 6, 275 

\bibitem[Villar-Mart{\'{\i}}n et al.(2003)]{villar03} Villar-Mart{\'{\i}}n M., Vernet J., di Serego Alighieri S., Fosbury R., Humphrey A., Pentericci L., 2003, \mnras,  346, 273 

\bibitem[Willott et al.(2002)]{willott02} Willott, C.~J., 
Rawlings, S., Archibald, E.~N., \& Dunlop, J.~S.\ 2002, \mnras, 331, 435 

\bibitem[Willott et al.(2003)]{willot03} Willott, C.~J., 
Rawlings, S., Jarvis, M.~J., \& Blundell, K.~M.\ 2003, \mnras, 339, 173 
 

\end{thebibliography}
\end{document}